\documentclass[10pt]{article}
\usepackage{amsmath,amsfonts,amssymb,amsthm,bbm}
\usepackage{pdfsync}
\usepackage{graphicx}
\usepackage[latin1]{inputenc}
\usepackage{amsthm}
\usepackage{hyperref}
\usepackage[all]{xy}
\usepackage{stmaryrd}

\newcommand{\Ref}[1]{(\ref{#1})}

\newcommand{\C}{{\mathbb C}}

\newcommand{\R}{{\mathbb R}}
\newcommand{\Z}{{\mathbb Z}}
\def\T{\mathbbm T}

\newcommand{\cC}{{\mathcal C}}
\newcommand{\cS}{{\mathcal S}}

\newcommand{\SU}{\mathrm{SU}}

\newcommand{\SL}{\mathrm{SL}}
\newcommand{\SO}{\mathrm{SO}}

\newcommand{\ISO}{\mathrm{ISO}}
\newcommand{\sltc}{\mathrm{SL}(2,\mathbb{C})}
\newcommand{\hh}{{\cal H}}

\newcommand{\be}{\begin{equation}}
\newcommand{\ee}{\end{equation}}
\newcommand{\bea}{\begin{eqnarray}}
\newcommand{\eea}{\end{eqnarray}}

\newcommand{\bit}{\begin{itemize}}
\newcommand{\eit}{\end{itemize}}

\newcommand{\w}{\wedge}

\newcommand{\tr}{{\rm Tr}}
\newcommand{\f}{\frac}
\newcommand{\tl}{\widetilde}
\def\p{\partial}
\newcommand{\Id}{\mathbbm{1}}

\newcommand{\re}{\mathrm{Re}}
\newcommand{\im}{\mathrm{Im}}
\newcommand{\ii}{\mathrm{i}}
\newcommand{\ex}{\mathrm{e}}
\newcommand{\dd}{\mathrm{d}}

\newcommand{\ra}{\rangle}

\newcommand{\bra}[1]{\langle {#1}|}
\newcommand{\ket}[1]{|{#1}\rangle}

\renewcommand{\a}{\alpha}  \newcommand{\g}{\gamma}
\renewcommand{\d}{\delta}  \newcommand{\eps}{\epsilon} \newcommand{\veps}{\varepsilon} \newcommand{\z}{\zeta}
 \renewcommand{\th}{\theta}      \renewcommand{\l}{\lambda}
     \newcommand{\s}{\sigma}       \let\om=\omega
\let\G=\Gamma   \let\Th=\Theta \let\L=\Lambda  

\newcommand{\po}{\pi\om}
\newcommand{\tpo}{\tl\pi\tl\om}
\newcommand{\tlop}{\tl\om\tl\pi}

\def\hN{\widehat{N}}
\def\cN{{\cal N}}
\newcommand{\wh}{\widehat}

\def\to{\widetilde{\omega}}

\usepackage[usenames,dvipsnames]{pstricks}
\usepackage{epsfig}
\usepackage{pst-grad} 
\usepackage{pst-plot} 

\begin{document}

\title{\bf Null twisted geometries}

\author{\Large{Simone Speziale and Mingyi Zhang}
\smallskip \\
\small{Centre de Physique Th\'{e}orique, CNRS-UMR 7332, Aix-Marseille Univ, Luminy Case 907, 13288 Marseille, France}}
\date{\empty}

\maketitle

\begin{abstract}
\noindent We define and investigate a quantization of null hypersurfaces in the context of loop quantum gravity on a fixed graph. 
The main tool we use is the parametrization of the theory in terms of twistors, which has already proved useful in discussing the interpretation of spin networks as the quantization of twisted geometries. 
The classical formalism can be extended in a natural way to null hypersurfaces, 
with the Euclidean polyhedra replaced by null polyhedra with spacelike faces, and SU(2) by the little group ISO(2).
The main difference is that the simplicity constraints present in the formalism are all first class, 
and the symplectic reduction selects only the helicity subgroup of the little group. 
As a consequence, information on the shapes of the polyhedra is lost, and the result is a much simpler, Abelian geometric picture.
It can be described by a Euclidean singular structure on the two-dimensional spacelike surface defined by a foliation of space-time by null hypersurfaces. This geometric structure is naturally decomposed into a conformal metric and scale factors, forming locally conjugate pairs. Proper action-angle variables on the gauge-invariant phase space are described by the eigenvectors of the Laplacian of the dual graph.
We also identify the variables of the phase space amenable to characterize the extrinsic geometry of the foliation.
Finally, we quantize the phase space and its algebra using Dirac's algorithm, obtaining a notion of spin networks for null hypersurfaces. Such spin networks are labeled by SO(2) quantum numbers, and are embedded nontrivially in the unitary, infinite-dimensional irreducible representations of the Lorentz group.

\end{abstract}

\section{Introduction}

Null hypersurfaces play a pivotal role in the physical understanding of general relativity. 
We are interested in understanding how null hypersurfaces can be described within loop quantum gravity (LQG), and their dynamical properties. Research in the dynamics of loop quantum gravity is mostly concerned with the evolution of spacelike hypersurfaces, in the spirit of the ADM (Arnowitt-Deser-Misner) canonical approach it is rooted on. It is commonly described by the spin foam formalism, or its embedding in group field theory. One considers transition amplitudes between fixed graphs, then refines or sums over the graphs. The boundary data assigned on the graphs are typically taken to be spacelike, however, the spin foam formalism is completely covariant, and in principle one can consider arbitrary configurations. Some results on timelike boundaries have appeared in \cite{AlexandrovKadar05, Conrady:2010kc}, but null configurations have received little attention so far.\footnote{For instance, a discussion of admissible null boundaries for spin foams has appeared in \cite{Neiman:2012fu}.}
To extend the description to null boundary data, the first step is to understand what null data mean from the viewpoint of LQG variables on a fixed graph. 
In this paper, we point out a natural answer suggested by the recent description of LQG in terms of twistors and twisted geometries \cite{twigeo,twigeo2,EteraHoloQT,IoPoly,WielandTwistors,IoHolo,IoTwistorNet,IoWolfgang,Freidel:2013fia}. 

Twistors describing LQG in real Ashtekar-Barbero variables satisfy a certain incidence relation \cite{IoWolfgang}, determined by the timelike vector used in the $3+1$ splitting of the gravitational action. Such constrained incidence relation is the twistor's version of the discretized (primary) simplicity constraints presenting in the Plebanski action for general relativity. The idea of this paper is to describe discrete null hypersurfaces by taking the vector appearing in the incidence relation to be null.
The first consequence of this choice is that the usual group $\SU(2)$ is replaced by $\ISO(2)$, the little group of a null vector. Furthermore, the primary simplicity constraints are all first class, and only the $\SO(2)$ helicity subgroup survives the  symplectic reduction: the translations are pure gauge. 
This fact has an appealing counterpart in particle theory: as well-known, the representations of massless particles only depend on the spin quantum number, the translations being redundant gauges. In our setting, the gauge orbits have the geometric interpretation of shifts along the null direction of the hypersurface.

In the next section, we briefly review polyhedra with spacelike faces in null hypersurfaces, and how they can be described in terms of bivectors satisfying the closure and simplicity constraints. In particular, we provide a gauge-invariant set of variables allowing us to reconstruct a unique null polyhedron starting from its bivectors. 
Because of the special isometries present due to the existence of null directions, such gauge-invariant variables are a little more subtle than the scalar products that one may immediately think of by analogy with the Euclidean case.
In Sec. 3, we describe the phase space of Lorentzian spin foam models with the null simplicity constraints and its description in terms of twistors, and show how the null polyhedra are endowed in this way with a symplectic structure. We then proceed to study the symplectic reduction, interpret geometrically the orbits of the simplicity constraints and identify the global isometries as well as the transformations changing the shapes of the polyhedra. The latter are also first class; thus the reduced phase describes only an equivalence class of null polyhedra, determined only by the areas and their time orientation.

The geometry of the two-dimensional spacelike surface  can be parametrized in purely gauge-invariant terms, and describes 
a Euclidean singular structure (see e.g. \cite{Carfora:2002rn}) with scale factors associated with the faces of the graph, instead of the nodes. These data are less than those characterizing a two-dimensional Regge geometry, again a peculiarity of the large amount of symmetry in the system. For planar graphs, the reduced Poisson brackets evaluate to the Laplacian matrix of the dual graph. Therefore proper gauge-invariant action-angle variables can be identified in terms of its eigenvectors. For nonplanar graphs the situation is slightly more complicated, as the matrix of Poisson brackets has off-diagonal elements of both signs.
Finally, we comment on the possible role played by secondary constraints that future studies of the dynamics may unveil, in particular, we identify the kinematical degrees of freedom amenable to describing the extrinsic geometry of the foliation.

In Sec. 5, we quantize the system and find an orthonormal basis for the reduced Hilbert space. Such null spin networks are labeled by $\SO(2)$ quantum numbers, and are naturally embedded in the lightlike basis of homogeneous functions used for the unitary, infinite-dimensional representations of the Lorentz group. The basis diagonalizes the oriented areas, and the (complex exponentials of the) deficit angles act as spin-creation operators. 
This paper is only a first, preliminary step toward understanding the dynamics of null surfaces in loop quantum gravity, 
and in the conclusions we comment on some next steps in the program, as well as desired applications.
Finally, the Appendix contains details and conventions on the Lorentz algebra and its $\ISO(2)$ subgroup.

\section{Simple bivectors and null polyhedra}

In this section, we describe how null polyhedra can be described in terms of bivectors.
By null polyhedra, we will mean polyhedra with spacelike faces living in a three-dimensional null hypersurface of Minkowski spacetime.
Consider a bivector $B^{IJ}$ in Minkowski spacetime, orthogonal to a given direction $N^I$,
\be\label{nB}
N_I B^{IJ} = 0.
\ee
The condition implies that the bivector is simple; namely it can be written in the form 
$B^{IJ} = 2u^{[I} v^{J]}$. 
The proof is straightforward, and valid for any signature of $N^I$.\footnote{An arbitrary bivector $B^{IJ}$ can be written as
$
B^{IJ}=a^{[I}b^{J]}-c^{[I}d^{J]}.
$
If \Ref{nB} holds, then
$\left( a\cdot N\right) b-\left( b\cdot N\right) a-\left(
c\cdot N\right) d+\left( d\cdot N\right) c=0$,
which implies that the four vectors are linearly dependent. Simplicity immediately follows, independent of the signature of $N^I$.} 
Provided $u$ and $v$ are linearly independent, the simple bivector identifies a plane, as well as a scale
$B^2:=B^{IJ} B_{IJ}/2$. 
When $N^I$ is null, the two vectors $u$ and $v$ can then be either null or spacelike. 
If they are both null, they both must be proportional to $N^I$, and thus the bivector is ``degenerate'' and does not span a plane.
In this paper we focus our attention on the case of spacelike bivectors. 

Such simple bivectors can always be parametrized as
\be\label{Bb}
B^{IJ} =\frac{1}{2} \eps^{IJ}_{~~KL} N^K b^L, \qquad b^2=0, \qquad B^2=(b\cdot N)^2.
\ee
We further denote $A:=|B|$, and $b\cdot N=-\veps A$, with $\veps=\pm$.

Next, take a collection of bivectors $B_l$, all lying in the same hypersurface determined by $N^I$, and further constrained by the closure condition 
\be\label{closB}
\sum_{l} B_l = 0.
\ee 
In the case of a timelike $N^I$, a theorem by Minkowski proves that the set defines a unique, convex and bounded polyhedron, with areas $A_l$ and dihedral angles determined by the scalar products among the bivectors. This fact plays a key role in the interpretation of loop quantum gravity in terms of twisted geometries. See \cite{IoPoly} for details and the explicit reconstruction procedure. An application of the same theorem to the case of null $N^I$ implies that the polyhedron now lies in the null hypersurface orthogonal to $N^I$, which includes $N^I$ itself. 
A null hypersurface has a degenerate induced metric, with signature $(0,+,+)$, and therefore the metric properties of the polyhedron are entirely determined by its projection on the spacelike 2d surface.\footnote{This does not mean that the null direction never plays a geometric role: it will acquire a geometrical meaning, if ones embeds the three-dimensional null hypersurface in a nondegenerate ambient space-time.}
In fact, one can arbitrarily translate the vertices of the polyhedron along the null direction without changing its intrinsic geometry. Using this symmetry, the polyhedron can always be ``squashed'' on the two-dimensional spacelike surface, where it will look like a degenerate case of a Euclidean polyhedron. It is indeed often helpful to visualize a null polyhedron as an ordinary polyhedron in coordinate space, endowed with a degenerate metric.

Using the parametrization \Ref{Bb} of simple bivectors, the closure condition can be equivalently rewritten as
\begin{equation}\label{eq:CloseC}
V^I := \sum_l b^{I}_{l} = \a N^I, \qquad \a\in \R.
\end{equation}
These are three independent equations, since $\a$ is arbitrary, and therefore the space of $F$ simple, closed bivectors has $3F-3$ dimensions. In particular, contracting both sides with $N_I$ we obtain the ``area closure'',
\be\label{areaClos}
-N\cdot V= \sum_l \veps_l A_l = 0. 
\ee
This condition is also satisfied by a degenerate Euclidean polyhedron squashed on a 2d plane, and it  allows us to identify $A_l$ with the areas of the null polyhedron's faces. Furthermore, assuming once and for all $N^I$ to be future pointing, and the normals outgoing to the faces, the sign $\veps_l$ measures whether the face $l$ is future or past pointing.
While \Ref{areaClos} plays a predominant role, one should not forget that the complete closure condition satisfied by the bivectors has two extra equations, contained in \Ref{closB} or \Ref{eq:CloseC}.
It is also interesting to note that \Ref{eq:CloseC} allows us to map the space of null polyhedra with $F$ faces to the space of null polygons with $F+1$ sides, with one direction held fixed, but we will not further pursue this interpretation here.

Another peculiarity of null polyhedra is to have a larger isometry group than their Euclidean brothers.
Clearly, global (i.e. acting on all bivectors) Lorentz transformations belonging to the little group of $N^I$, which is the Lie group $\ISO(2)$, do not affect the intrinsic geometry. But there is an additional isometry due to the degeneracy of the induced metric: boosts along the $N^I$ direction do not change the intrinsic geometry of the polyhedron, because the induced metric is degenerate along that direction. Therefore, the isometry group has four dimensions, and the space of shapes of null polyhedra has $3F-7$ dimensions.

\begin{figure}[!htbp]
\centering
\includegraphics[width=0.3\textwidth]{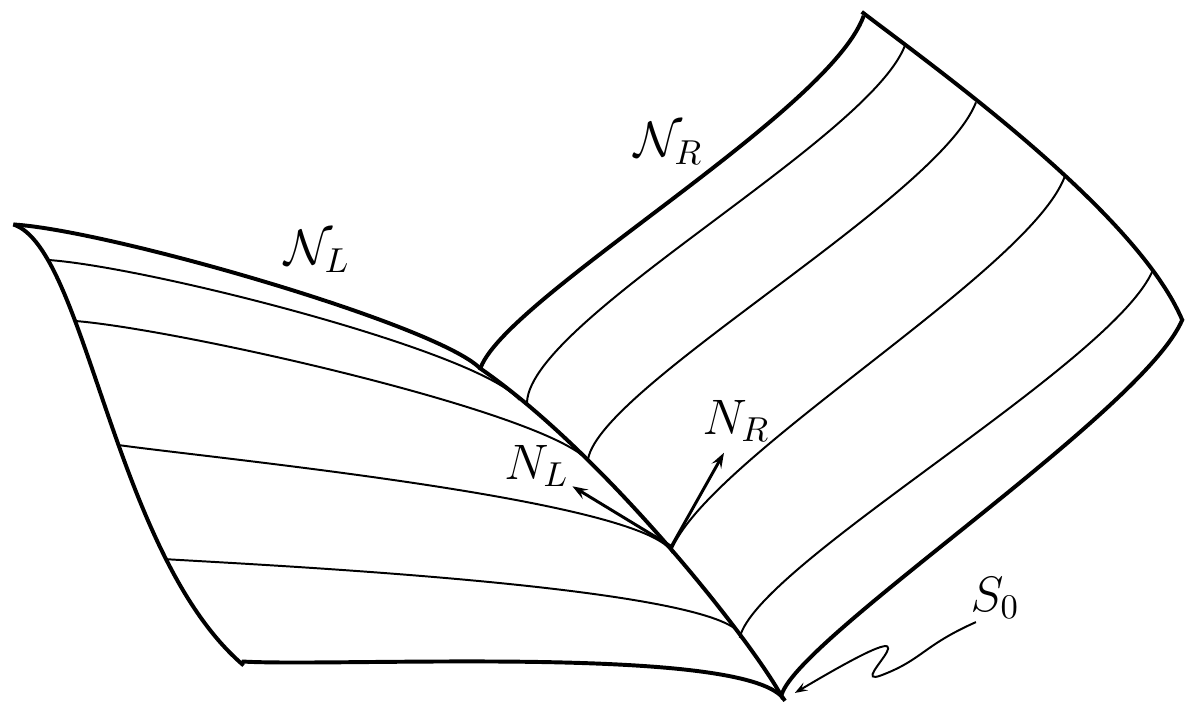}
\caption{A foliation of spacetime by null hypersurfaces.}
\label{FigFoliation}
\end{figure}

An interesting question is how to parametrize the intrinsic shapes of null polyhedra. 
In the Euclidean case, we are used to do so using the scalar products between the normals within the hypersurface, which fully respect the isometries. However, this is not the case for null polyhedra, where it is the common normal $N^I$ to lie in the hypersurface, while the null normals $b^I_l$ characterizing the individual faces do not lie in the hypersurface, and need not respect the isometries.
For instance, translating a vertex of the polyhedron along the null direction is an isometry, but this transformation does not preserve the scalar product between the null normals $b^I_l$. 
Conversely, while individual simple bivectors define planes, the intersection of planes cannot be defined in a degenerate metric. 
Therefore, the characterization of the intrinsic shapes cannot be done solely in terms of the $b_l$; one must resort to the full Minkowski spacetime and its nondegenerate metric. 
To fix ideas, consider the foliation of Minkowski spacetime generated by $\cN$ and $\wh\cN$, the null hypersurfaces defined , respectively by $N^I$ and its parity transformed $\wh N^I={\cal P}N^I$, satisfying $\wh N\cdot N=-1$. See Fig. \ref{FigFoliation}.

Using both normals, one can make sense of the intersection of two faces, say $l$ and $l'$, within $\cN$, and characterize it by the (pseudo)vector
\be\label{mighty}
\tl E^I_{ll'} = \eps^{IJKL} N_J (\epsilon_{KMPQ} \hN^M B_l^{PQ}) ( \epsilon_{LRST} \hN^R B_{l'}^{ST}).
\ee
With this formula, one can explicitly reconstruct the intrinsic shape of the null polyhedron starting from the bivectors. To show this, let us first consider the case of a tetrahedron, and then a general polyhedron.

The simplicity of the tetrahedral case lies in its trivial adjacency matrix: any two faces identify an edge of the tetrahedron, and the intrinsic shapes can be described by any three edge vectors meeting at one vertex, by providing the lengths and the angles among them. The existence of a null direction will show up explicitly in the fact that only two of the angles are linearly independent, thus the intrinsic shape is characterized by only five quantities.
Consider then three faces, say $l=1,2,3$, and the three edges determined by their intersections.
Let us first assume that the three edge vectors are not coplanar in $\cal N$ (the degenerate case will be dealt with later). Then, we define
\be
V_c(B)^4 := - \f1{6^4} \eps_{IJKL} \hN^I \tl E_{13}^{J}(B) \tl E_{21}^{K}(B) \tl E_{32}^L(B).
\ee
The right-hand side is always positive, and defines a coordinate volume of the tetrahedron, analogous to the definition of the Euclidean volume in terms of the triple product. We can then normalize \Ref{mighty} and obtain the proper edge vectors of the tetrahedron as
\be\label{ENbb}
E^I_{ll'} := \f1{6 V_c} \tl E^I_{ll'} = -\f1{6V_c} \eps^I{}_{JKL} N^J b_{l}^K b_{l'}^L,
\ee
where we used \Ref{Bb}.
Finally, the edge lengths and angles of the triple evaluate to
\begin{subequations}\label{lengths}\begin{align}
& E^2_{ll'} = -\f2{(6V_c)^2} \, (b_{l} \cdot N) \, (b_{l'} \cdot N) \, (b_{l} \cdot \, b_{l'}), \\
& E_{ll'} \cdot E_{l'l''} =  \f1{(6V_c)^2} \, \Big[(b_{l} \cdot N) \, (b_{l'} \cdot N) \, (b_{l'} \cdot \, b_{l''})
+ (b_{l'} \cdot N) \, (b_{l''} \cdot N) \, (b_{l} \cdot \, b_{l'}) + (b_{l'} \cdot N)^2 \, (b_{l} \cdot \, b_{l''}) \Big]. \label{angles}
\end{align}\end{subequations}
It is easy to check that we can always consistently pick $B^{IJ}_l = 2E_{ll'}^{[I} E_{l''l}^{J]}$, and that the triangles' areas computed from the edge vectors coincide with $A_l$. 
Furthermore, the oriented sum of the angles defined by \Ref{angles} vanishes, so that only five quantities out of the six defined in \Ref{lengths} are independent.

The formulas \Ref{lengths} provide the intrinsic shape of the null tetrahedron in terms of simple bivectors. They are valid for any time orientation of the faces and, as promised, are left invariant when any of the vectors is  translated along the null direction $N^I$. In particular, this makes the expressions for edges and angles valid also in the special case when the isometry is used to ``squash'' the tetrahedron down to the spacelike surface $S_0$. When this happens, the $b_l^I$ are all parallel, so their scalar products vanish, but also $V_c$ vanishes, and the ratio $(b_{l} \cdot \, b_{l'})/V_c^2$ remains finite. Hence \Ref{lengths} are well defined also in the limit case when the edge vectors are coplanar.
We conclude that the intrinsic geometry can be characterized in terms of the null vectors $b_l^I$, using the scalar products $b_l\cdot N$ as well as the ratios $(b_{l} \cdot \, b_{l'})/V_c^2$,
of which only two out of three are independent.
On the other hand, notice that the scalar products $b_l \cdot b_m$ are not good variables: they are not preserved by the isometries, and different values can correspond to the same intrinsic geometry.

The main difficulty to extend this construction to higher polyhedra comes from the fact that the adjacency matrix is not trivial anymore: the explicit values of the bivectors themselves will determine whether two faces are adjacent or not.
A strategy to deal with this case is to use the reconstruction algorithm already developed for the Euclidean signature.
To that end, we work in light-cone coordinates defined by $N^I$ and $\hN^I$. 
In these coordinates, the closure constraint \Ref{Gn} identifies a closure condition for 3d vectors in a space with a degenerate metric of signature $(0,+,+)$. If we replace this metric by an auxiliary Euclidean metric, we can apply the reconstruction procedure of \cite{IoPoly} to the resulting Euclidean polyhedron. In particular, compute its adjacency matrix, and once this is known, apply \Ref{lengths} to the existing edges to determine the null geometry of the polyhedron.
It would be interesting to know whether the adjacency matrix of a null polyhedron can be reconstructed directly from the $b_l^I$, without passing through the auxiliary Euclidean reconstruction, but this is not needed for the rest of the paper, and we leave it as an open question.

Finally, recall that 
the space of shapes of 3d Euclidean polyhedra has dimensions $3F-6$, and the $2F-6$ space of shapes at fixed areas is a phase space \cite{Kapovich}, a result used in the twisted geometry parametrization \cite{IoPoly}.
This turns out not to be the case for null polyhedra, because as we show below, the closure condition does not generate all the isometries. 
While it is an interesting open question to construct a phase space of shapes for null polyhedra, 
we will see below that the phase space of loop gravity on a null hypersurface does include a description of polyhedra, but rather as equivalence classes, defined by their areas only.

\section{Null simplicity constraints in LQG}

Spin foams are based on the nonchiral Plebanski action for general relativity, 
\be\label{action}
S(\om^{IJ},B,\psi) = \int \tr \Big(\star +\f\Id\g \Big) B\w F(\om^{IJ}) + \psi_{IJKL} B^{IJ}\w B^{KL},
\ee
where the fundamental variables are a Lorentz connection $\om_\mu^{IJ}$, and a 2-form valued in the Lorentz algebra $B^{IJ}$, constrained by $\psi_{IJKL}$ to be simple, that is $B^{IJ}=e^I\w e^J$. Here $\g$ is the Immirzi parameter, and we assumed a vanishing cosmological constant. 
The canonical analysis of this action has been studied in a number of papers (e.g. \cite{AlexandrovBuffenoir}), and we refer the reader to the living review \cite{PerezLR} for details and an introduction to the spin foam formalism. The phase space is described by the pullback of the Lorentz connection and its conjugate momentum, that is the pullback of the 2-form 
\be\label{MtoB}
M^{IJ}=\Big(\star +\f\Id\g \Big) B^{IJ}, \qquad
B^{IJ} =\f{\g}{\g^{2}+1}\Big( \Id- \g \star\Big) M^{IJ}.
\end{equation}

In the following, we are interested in a discretized version of this canonical structure, which is commonly used in the construction of spin foam models \cite{PerezLR}.
The discrete variables are distributional smearings along an oriented graph $\G$, say with $L$ links and $N$ nodes, where the gravitational connection is  replaced by holonomies $h_l$ along the links, and the conjugate momentum by algebra elements $M_l$, referred to as fluxes. The phase space associated with a graph is
\be\label{PG}
 P_\G = T^*\SL(2,\C)^L, \qquad (M_l, h_l) \in T^*\SL(2,\C),
\ee
which notably comes with a noncommutativity of the fluxes. 
This kinematical phase space appears in Lorentzian spin foam models \cite{EPRL}, as well as in covariant loop quantum gravity \cite{AlexandrovLivine}. We then consider two sets of constraints on the $B$ variables. The first is a discrete Gauss law, or closure condition, 
\be\label{Gn}
G_n^{IJ} = \sum_{l\in n} B_l^{IJ} = 0.
\ee
It is local on the nodes of the graph, and it imposes gauge invariance. The second is a discrete version of the simplicity constraints,
\be\label{linsimpl}
S^J_{nl} = N_{nI} B_l^{IJ} = 0, \qquad \forall l\in n,
\ee
where $N^I_n$ is a unit vector assigned independently to each node $n$. This linear version of the discrete simplicity constraints was introduced in \cite{EPR}, with $N^I$ timelike and related to the hypersurface normal used in the $3+1$ decomposition of the action.
We denote ${\cal S}_\G$ the reduced phase space obtained imposing the constraints \Ref{Gn} and \Ref{linsimpl},
\be\label{SG}
{\cal S}_\G = T^*\SL(2,\C)^L/\!/F_{nl}/\!/G_n.
\ee

When $N^I$ is timelike, it was shown in \cite{IoWolfgang} that ${\cal S}_\G\equiv T^*\SU(2)^L/\!/\SU(2)^N$, where for any finite $\g\neq 0$, the relevant $\SU(2)$ subgroup is not the canonical subgroup of the Lorentz group, but a group manifold nontrivially embedded in $T^*\SL(2,\C)$, capable in particular of probing boosts degree of freedom. The interpretation of $\cS_\G$ is that of a truncation of general relativity to a finite number of degrees of freedom \cite{IoCarloGraph}, whose geometry can be described  by twisted geometries \cite{twigeo}. 

In this paper we investigate the consequences of taking vector $N^I$ in \Ref{linsimpl} to be null, and derive a geometric description for the reduced space \Ref{SG}, in the spirit of twisted geometries.
Ideally, this should be related to a formulation of the Plebanski action in which we perform a standard $3+1$ splitting, and use the internal Minkowski space to induce a noninvertible 3d metric with signature $(0++)$. 
The continuum canonical analysis of \Ref{action} in this null setup, as well as studying the resulting dynamical structure, will be investigated elsewhere.\footnote{In particular, the analysis is expected to reveal the presence of secondary constraints, which should play an important role in the identification of the extrinsic geometry, as we will discuss below.}
Our goal here is simply to study \Ref{SG} when $N^2=0$, its geometrical interpretation, and its quantization.

We will proceed in two steps, motivated by the structure of \Ref{SG}. First, we focus on a single link, studying the phase space $T^*\SL(2,\C)$ and the pair of simplicity constraints \Ref{linsimpl}, which are local on the links. At a second stage, we consider the full graph structure and the closure condition \Ref{Gn}.

\subsection{Phase space structure}

We saw in Sec. 1 that a set of bivectors satisfying closure and simplicity defines polyhedra. 
The polyhedra can be endowed with the symplectic structure of $T^*\SL(2,\C)$ via \Ref{MtoB} and \Ref{PG}, as follows. 
Picking a specific time direction $t^I=(1,0,0,0)$, we identify boosts, rotations and chiral left-handed generators, respectively, as 
$$K^i:=M^{0i}, \qquad L^i=-\f12\eps^i{}_{jk} M^{jk}, \qquad \Pi^i = \frac{1}{2}(L^i+\ii K^i) = \ii \s^{iA}{}_{B} \Pi^{B}{}_A.$$
Here  $A,B=0,1$ are  spinorial indices, raised and lowered with the antisymmetric symbol $\eps^{AB}$, and $\s^{A}{}_B$ the Pauli matrices. See Appendix for a complete list of conventions, notations and background material.
We parametrize $T^*\SL(2,\C)$ via the pair $(\Pi^{A}{}_{B}, h^{A}{}_B)$, with $h$ a group element in the fundamental $({\bf 1/2,0})$ representation, and symplectic potential 
$\Th= \tr(\Pi hd h)+cc$. The $\Pi$ are left-invariant vector fields, and $\widetilde\Pi=-h\Pi h^{-1}$ right-invariant ones.
We can equivalently use the parametrization $(\Pi, \tl \Pi)$ and the complex angle Tr$(h)$. In this way, we can associate a generator, and thus a bivector $B$ through \Ref{MtoB}, with both source and target nodes of a link. Hence, we can consider the topological polyhedra defined by a cellular decomposition dual to the graph, and associate a bivector $B$ with each face within each frame. By construction, a face inherits two bivectors, and unique norm, $B^2=\tl B^2$, and we notice that the closure condition \Ref{Gn} is equivalent to closure for the generators. 

The simplicity conditions \Ref{nB} introduce a preferred direction via $N^I$, thus reducing the initial Lorentz symmetry to its little group. For a null vector, the Lie group $\ISO(2)$.
To fix ideas, we take from now on the specific null vector $N^I=(1,0,0,1)/\sqrt{2}$, with the normalization chosen for later convenience. Its little group $\ISO(2)$ is generated by 
$$ L^3, \qquad P^1:=L^{1}-K^{2}, \qquad P^2:=L^{2}+K^{1},$$ 
and the simplicity constraints \Ref{linsimpl} read
\begin{equation}\label{nullsimplGen}
\gamma L^{3}+K^{3}=0, \qquad P^a=0, \qquad a=1,2.
\end{equation}

There are two important differences with respect to the timelike case. First of all, the constraints impose the vanishing of part of the little group itself, thus effectively selecting its helicity $\SO(2)$ subgroup. Second, by themselves they form a completely first class system, unlike in the timelike case, as can be verified trivially. 
These facts have important consequences for the geometric interpretation of the reduced phase space. 
To study the symplectic reduction and its geometric interpretation, we use the twistorial parametrization introduced and studied in \cite{twigeo2,WielandTwistors,IoHolo,IoTwistorNet,IoWolfgang}.

\subsection{Twistorial description}

A twistor can be described as a pair of spinors,\footnote{The presence of an $i$ differs from the standard Penrose notation, and it is just a matter of convenience to bridge with the conventions used in loop quantum gravity.} 
$Z^\a=(\omega^A,i \bar\pi_{\dot A})\in\mathbb{C}^2\oplus\bar{\mathbb{C}}^2{}^\ast=:\mathbb{T}$.
The space then carries a representation of the Lorentz algebra, which preserves the complex bilinear
$\pi_A\om^A\equiv \po$.
To describe the symplectic manifold $T^*\SL(2,\C)$ on an oriented link, we consider a pair $(Z, \tl Z)$ associated , respectively,  with the source and target nodes of the link, and equip each twistor with canonical Poisson brackets,
\begin{equation}\label{eq:CPoiB}
  \{\pi_A,\omega^B\}=\delta_A^B=\{\widetilde\pi_A,\widetilde\omega^B\}.
\end{equation}
We then impose the following area-matching condition,
\be\label{defC}
C= \po - \tlop = 0.  
\ee
This is a first class complex constraint generating the scale transformations
$
(\omega,\pi,\widetilde\omega,\widetilde\pi)\mapsto(\ex^{z}\omega, \ex^{-z}\pi, \ex^{z}\widetilde\omega,\ex^{-z}\widetilde\pi).
$
The 12d manifold obtained by symplectic reduction by \Ref{defC} coincides with $T^*\SL(2,\C)$, with holonomies and fluxes that can be parametrized as
\begin{equation}\label{hfpar}
\Pi^{AB} = \f{1}{2}\om^{(A}\pi^{B)}, \qquad  h^A{}_B=\f{\widetilde\om^A\pi_B+\widetilde\pi^A\om_B}{\sqrt{\po}\sqrt{\tlop}},
\end{equation}
and 
\be\label{PitildePi}
\tl \Pi^A{}_B = \f{1}{2}\tl\om^{(A}\tl\pi^{B)} \equiv -h^A{}_C\Pi^C{}_D h^{-1}{}^D{}_B. 
\ee
As it is apparent from \Ref{hfpar}, the parametrization is valid provided $\po$ and $\tpo$ do not vanish. The submanifold where this occurs can be safely excluded: it would correspond to null bivectors, whereas we are restricting attention to spacelike bivectors. Notice also that the parametrization is 2-to-1, as it is invariant under the exchange of spinors,
\be\label{Z2sym}
(\om,\pi,\tl\om,\tl\pi) \mapsto (\pi,\om,\tl\pi,\tl\om).
\ee
See \cite{IoWolfgang} for further details.\footnote{Note however that the conventions here are slightly different. This change, consistent with other upcoming papers \cite{IoMiklos,IoFabio}, is motivated by the desire of having the same Poisson brackets for source and target twistors. The burden of keeping track of the link orientation is put on the holonomy, which transforms the basis with a minus sign, $h \om = \tl\om, \ h\pi = -\tl\pi.$ This conveniently ``flips'' the orientation of the $\C^2$ basis in a way consistent with the usual convention of orienting all normals as locally outgoing.}
To write the simplicity constraints, we introduce a canonical basis in $\C^2$, $(o^A=\d^A_0, \iota^A=\d^A_1)$. The chosen null vector reads $N^{A\dot A} = i o^{A}\bar{o}^{\dot{A}}$, and \Ref{nB} becomes
\begin{equation}\label{Pisimpl}
N_{A \dot{A}} \Pi^{AB}\eps^{\dot{A}\dot{B}} = \ex^{i\theta }N_{A \dot{A}} \eps^{AB}\bar\Pi^{\dot{A}\dot{B}},
\qquad \ex^{\ii\theta}\equiv(\gamma+\ii)/(\gamma-\ii).
\end{equation}
Notice that the matrix $\d^o{}^{A\dot A}:= o^{A}\bar{o}^{\dot{A}}$ defines an Hermitian scalar product, $|\!|\om|\!|^2 = |\om^1|^2$, preserved by the little group $\ISO(2)$. 
The above conditions can be conveniently separated as
\be\label{Fdef}
F_1 = \re (\po) - \g \, \im(\po) = 0, \qquad F_2 = o_A\bar o_{\dot A} \om^A \bar\pi^{\dot A} = \om^1 \bar\pi^1=0,
\ee
where $F_1$ is real and Lorentz invariant, whereas $F_2$ is complex and only $\ISO(2)$ invariant. 
In particular, $F_2$ imposes $P^a=0$, and on-shell of this condition $F_1$ reduces to the first condition in \Ref{nullsimplGen}.
The structure is very similar to the timelike case of \cite{IoWolfgang}: in particular, the Lorentz-invariant part $F_1$ is the same, and
can be solved posing
\be\label{gij}
\po = (\g+i)\varepsilon j, \qquad \varepsilon=\pm, \qquad j\in\R^+.
\ee
With this parametrization, $\veps$ determines the sign of the twistor's helicity: $\veps=+$ for positive helicity.
Notice that the $\Z_2$ symmetry \Ref{Z2sym} of the twistorial parametrization flips this sign, therefore it is possible to fix $\varepsilon=1$ without loss of generality in parametrizing $T^*\SL(2,\C)$. 
$F_2=0$ has two solutions, $\om^1=0$ and $\pi^1=0$. Both branches are needed to describe the reduced phase space, introducing a slightly awkward notation, where the reduced phase space is parametrized partly by $\om^A$ and partly by $\pi^A$. It is convenient to avoid this by exploiting the $\Z_2$ symmetry, since \Ref{Z2sym} switches between the two branches. It then turns out to be convenient to keep the $\varepsilon$ sign in \Ref{gij} free, and pick a single branch of $F_2=0$.
Let us assume $\om^1\neq 0$, and pick the solution $\pi^1=0$. 

The five-dimensional surface of \emph{simple} twistor solutions of \Ref{Fdef} can be parametrized by $(\om^A, j)$,
and
\be\label{pi=rom}
\pi^A = - r e^{i\f\th2} \d^o{}^{A\dot A} \bar\om_{\dot A}, \qquad r = \f{\veps j \sqrt{1+\g^2} }{|\!|\om |\!|^2}.
\ee
On this surface, the simplicity constraints generate the following gauge transformations,
\be
\{ F_1, \om^A \} = \f{1+\ii\gamma}{2}\om^A, \qquad \{F_2, \om^A \} = 0,   \qquad \{\bar F_2, \om^A \} = -\d^{A}_0 \bar\om^1,
\qquad \{F_1, j\} = \{F_2,j\}=0.
\ee
For the nontivial ones, the finite action is
\be\label{orbitsF}
e^{\{\a F_1, \cdot\} }\om^A = \ex^{ \f{1+\ii\gamma}{2}\a }\om^A, \qquad
e^{\{\a \bar F_2, \cdot\} }\om^A =  \om^A - \a \d^{A}_0 \bar\om^1.
\ee
We see that $\om^0$ is pure gauge and that $\om^1$ contains a dependence on the  gauge generated by $F_1$.
The gauge invariant reduced space has two dimensions, and can be parametrized by the following complex variable,
\be\label{defz}
z= \f{\sqrt{2j}}{|\!|\om |\!|^{i\g+1}} \om^1, \qquad |z|^2 = 2j,
\ee
plus the sign $\veps$. Notice that shifting the phase of $z$ by $\pi$ has the same effect as switching the sign of $\veps$. Hence, with our choice of parametrization $\arg(z)\in[0,\pi)$, to avoid covering twice the same space. In this way we identify the positive complex half-plane with positive helicities, and the negative half-plane with negative helicities.
The reduced symplectic potential evaluates to
\be\label{PBz}
\Th_{\rm red} = -\f i2 \veps zd\bar z + cc, \qquad \{z, \bar z\} = i \veps,
\ee
so the sign of the helicity determines the sign of the Poisson brackets. In conclusion, the symplectic reduction gives
$\T/\!/F = T^*S^1$, with the circle parametrized by two half-circles via $\arg(z)\in[o,\pi), \veps=\pm$.

To better understand the geometric meaning of the orbits of the simplicity constraints, it is useful to look at the bivectors $B^{IJ}$. These are given by \Ref{MtoB} in terms of the algebra generators $M^{IJ}$, whose spinorial form reads, from \Ref{hfpar}, $M^{IJ} = -\om^{(A} \pi^{B)} \eps^{\dot A\dot B} +cc$.
Introducing the following doubly null reference frame, 
\be
\ell^I = i \om^A\bar\om^{\dot A}, \quad k^I = i \pi^A\bar\pi^{\dot A}, \quad
m^I = i \om^A\bar\pi^{\dot A}, \quad \bar m^I = i \pi^A\bar\om^{\dot A}, \quad \ell\cdot k= - |\po |^2 = - m\cdot\bar m,
\ee
we can rewrite the bivectors as
\be\label{Bmm}
B^{IJ} = \f\g{1+\g^2} \f{2}{|\po|^2} \left[ (\g I-R) \ell^{[I} k^{J]} + \ii(\g R+ I) m^{[I} \bar m^{J]} \right]
\approx \f{2\ii \varepsilon\g}{j(1+\g^2)} m^{[I} \bar m^{J]},
\ee
where $\approx$ means that the equality holds on the constraint surface. The last equation defines a spacelike plane, and a scale $B^2=\g^2 j^2$, which represent the spacelike projection of the polyhedron's face. 
Comparing \Ref{Bmm} and \Ref{Bb}, we derive a parametrization of the normal null vector
$b^I$ in terms of spinors,
\be\label{b}
b^I = \f{\varepsilon\g j}{\|\om\|^2} \ell^I, \qquad b\cdot N = -\eps\g j.
\ee
Hence, we can also identify the helicity sign in \Ref{gij} with the sign of the time component of the face normal in \Ref{areaClos},
and since we are doing this identification for the ``untilded'' variables, it means that it holds provided the link is oriented outgoing from the node. 

It is straightforward to see that the orbits of $F_1$ leave the bivector $B^{IJ}$ as well as $b^I$ invariant. On the other hand, $F_2$ changes $b^I$, and its action can be used to always align this null vector with $\wh N^I=1/\sqrt{2}(1,0,0,-1)$. Hence, the orbits of $F_2$ allow us to project the face on the spacelike surface $S_0$ orthogonal to both $N^I$ and $\wh N^I$. This action becomes even clearer if we look at the spacelike vectors spanning the triangle,
\begin{subequations}\label{Totti}\begin{align}
& e^{\{-\a \bar F_2-\bar\alpha F_2, \cdot\} } \re(m)^I \approx \re(m)^I + \varepsilon j [\gamma \re(\alpha)+\im (\alpha)]N^I, \\
& e^{\{-\a \bar F_2-\bar\alpha F_2, \cdot\} } \im(m)^I \approx \im(m)^I + \varepsilon j [\re(\alpha)- \g \im (\alpha)]N^I.
\end{align}\end{subequations}
If we do this globally on all links around a node, that is we take $\a_l\equiv\a, \, \forall l$, we obtain the isometry corresponding to shifting the vectors along the null direction, and this action can be used to project all the faces to $S_0$.
On the other hand, acting independently on each link will genuinely deform the polyhedron, and can in principle break it open.
We will come back to this important point below in Sec. 4.
The geometric meaning of the action of $F_1$ will become clear next, when we discuss the reduction on the holonomy.

Let us conclude this section with a side comment, on the exact relation between the null simplicity constraints, and the usual twistor incidence relation. To that end, it is more convenient to look at the other solution of $F_2=0$, that is $\om^1=0$. 
This solution is equivalent to the one $\pi^1=0$ in the sense that this solution can be obtained from the $\Z_2$ symmetry \ref{Z2sym}. In this case, the simplicity conditions can then be packaged as the following constrained incidence relation,
\be\label{incidence}
\om^A = iX^{A\dot A} \bar\pi^\g_{\dot A},  \qquad 
X^{A\dot A} = -\f{\veps j \sqrt{1+\g^2}}{|\!|\pi|\!|^2}  n^{A\dot A}, \qquad \bar\pi^\g_{\dot A} = e^{i\f\th2}\bar\pi_{\dot A}.
\ee
From the point of view of twistor theory, \Ref{incidence} implies that (i) the twistor is $\g$-null \cite{IoMiklos}, namely that it is isomorphic to a null twistor, the $\g$-dependent isomorphism being $(\om, \pi)\mapsto (\om, \pi^\g:= e^{-i\th/2}\pi)$; and that (ii) the null ray $X^{A\dot A}$ described by the associated null twistor is aligned with $n^I$ and ``truncated'':
a simple twistor describes a specific null vector, and not anymore a null ray.

\subsection{Symplectic reduction, $T^*\ISO(2)$ and $T^*\SO(2)$}

To study the symplectic reduction on the link phase space, we consider two twistors $Z$ and $\tl Z$,
and impose the simplicity constraints \Ref{Fdef} on both, in agreement with \Ref{linsimpl}, as well as the area-matching condition \Ref{defC}. The complete system is first class, and partially redundant: $C=0=F_1$ implies $\tl F_1=0$. 
The simplicity constraints in the ``tilded'' sector can be solved in the same way,
\begin{equation}\label{tpisol}
\tl\pi^A = - \tl r e^{i\f\th2} \d^o{}^{A\dot A} \bar{\tl\om}_{\dot A}, \qquad \tl r = \f{\tl\veps \tl \jmath \sqrt{1+\g^2} }{|\!|\tl\om |\!|^2}.
\end{equation}
The area matching \Ref{defC} then imposes $\tl \veps \tl \jmath= -\veps j$, which we solve fixing $\tl \jmath = j$ and $\tl\veps=-\veps$. The opposite sign between $\veps$ and $\tl\veps$ keeps track of the sign difference between $\Pi$ and $\tl\Pi$ in \Ref{PitildePi}. 
As a consequence, a face which is future pointing in the frame of the source node is past pointing in the frame of the target node: following the same steps leading to \Ref{b}, we find $\tl b\cdot\tl N = -\tl\veps \g j=\veps\g j.$ In other words, $\veps$ coincides with the time orientation in the frame of the source node, and with its opposite in the frame of the target node.

On the seven-dimensional surface $\cC \subset T^*\SL(2,\C)$, where the simplicity constraints hold, fluxes and holonomies are
\begin{subequations}\label{redhf}\begin{align}
& \Pi^{A}{}_B \approx \f{(\g+i)\veps j}4 \left(\begin{array}{cc} -1 & 2{\om^0}/{\om^1} \\ 0 & 1 \end{array}\right), \qquad 
\tl \Pi ^{A}{}_B \approx - \f{(\g+i)\veps j}4 \left(\begin{array}{cc} -1 & 2{\tl\om^0}/{\tl\om^1} \\ 0 & 1 \end{array}\right), \qquad 
\\ & h^A{}_B \approx \left(\begin{array}{cc} {\om^1}/{\tl\om^1} & {\tl\om^0}/{\om^1}-{\om^0}/{\tl\om^1} \\ 
0 & {\tl\om^1}/{\om^1} \end{array}\right). \label{hred}
\end{align}\end{subequations}
As expected, the generators are restricted to those of the little group (up to the phase introduced by the Immirzi angle). 
The group element is also restricted, to a form which includes the little group $\ISO(2)$ as well as the extra isometry generated by a boost along the null direction ($K_3$ with our gauge choice for $N^I$). We can conveniently parametrize it as
\be\label{defu}
h \approx e^{\f12 \Xi \s_3} \ u, \qquad u = e^{\f12 \Xi \s_3} \ e^{-\ii \f12(\xi-\g \Xi) \s_3} \, T(\om^0,\tl\om^0) \ \in \ \ISO(2),
\ee 
where the boost rapidity is 
\be
\Xi:= \ln \f{|\!|\om |\!|^2}{|\!|\tl\om |\!|^2}, 
\ee
and we also defined
\be
\xi:= -2 \arg(z) -2\arg(\tl z) \in [0,4\pi).
\ee
Finally, the translational part 
\be
T(\om^0,\tl\om^0) = \left(\begin{array}{cc} 1 & {\tl\om^0}/{\om^1}-{\om^0}/{\tl\om^1} \\ 0 & 1 \end{array}\right)
\ee
vanishes when $\om^0$ and $\tl\om^0$ do, a fact that plays an important role below. 

A key aspect of this result is that the boost rapidity $\Xi$ enters also the rotational part of $h$. This is a consequence of the mixing between rotations and boosts introduced by the Immirzi parameter [see  \Ref{MtoB}], and it is presented also in the timelike case \cite{IoWolfgang}: it is the discrete equivalent of the mixing in the real Ashtekar-Barbero connection defined by 
$A^i_{a}= \om^i_a + (\g-\ii)K^i_a$, where $\om^i_a$ is the anti-self-dual part of the Lorentz connection and $K^i_a$ the (triad projection of the) extrinsic curvature. Loosely speaking, the mixing allows us to probe the Lorentzian phase space through a smaller subgroup, $\SU(2)$ in the timelike case and $\ISO(2)$ here. But while in the timelike case the holonomy on the constraint surface
is still a generic $\SL(2,\C)$ element \cite{IoWolfgang}, in the present null case it is a restricted group element, missing the algebra directions $\wh P^a$ capable of changing the direction of the vector $N^I$, a fact whose consequences will show up below.
Concerning the Poissonian structure of $\cC$, the symplectic potential of $T^*\SL(2,\C)$ restricted by the simplicity constraints  contains a piece generating the canonical Poisson brackets of $T^*\ISO(2)$ between $\Pi$ and $u$, and a degenerate direction. Therefore, $\cC$ contains a proper symplectic submanifold, and can be identified at least locally with the Cartesian product $T^*\ISO(2)\times \R$, where the additional dimension corresponds to boosts along $N^I$. The cotangent bundle of the little group thus appears at the level of the constraint surface. However,  a good part of it is just gauge, as we now show.

The next stage of the symplectic reduction is to divide by the gauge orbits. 
The gauge orbits of $F_1$ and $F_2$ have been studied in the previous sections: they amount to linear shifts of 
$\|\om\|$ and $\om^0$ , respectively. The latter are thus good coordinates along the orbits, and the gauge invariant part is the complex variable $z$ introduced in \Ref{defz}. The situation is analogous for the tilded variables, corresponding to the twistor associated with the second half of the link. In this case, we parametrize the reduced variable as
\be\label{deftlz}
\bar{\tl z} = \f{\sqrt{2j}}{|\!|\tl\om |\!|^{i\g+1}} \tl\om^1, \qquad |\tl z|^2 = 2j, \qquad  \{\tl z, \bar {\tl z}\} = i \veps.
\ee
Notice the extra complex conjugation appearing here, a convention taken to preserve the same sign of the brackets of $\tl z$ as for $z$. 
Proceeding in this way we have reduced by both $F_1$ and $\tl F_1$, and thus by part of the area-matching constraint \Ref{defC}. The remaining part is $C_{\rm red}:= |z|^2-|\tl z|^2=0$, which is already satisfied by the fact that we took in \Ref{deftlz} the same $j$ as in \Ref{defz}.
Its gauge transformations generate opposite phase shifts,
\be
\{C_{\rm red}, \arg(z)\} = -\veps = -  \{C_{\rm red}, \arg(\tl z)\}.
\ee
Hence, $\arg(z)-\arg(\tl z)$ is a good coordinate along the orbits, and $\xi=-2\arg(z)-2\arg(\tl z)$ previously defined is gauge invariant.
The two-dimensional reduced phase space on a link is thus spanned by the pair $(\veps j,\xi)$, which turns out to be canonical,
\be
\{\veps j,\xi\}=1.
\ee

Eliminating the gauges from \Ref{redhf}, we see that the reduced link phase space coincides with $T^*\SO(2)$,
\be\label{SO2hf}
X^{A}{}_B = \f{(\g+i)\veps j}4 \left(\begin{array}{cc} -1 & 0 \\ 0 & 1 \end{array}\right), \qquad 
g^A{}_B = \left(\begin{array}{cc} e^{-i\xi/2} & 0 \\ 
0 & e^{i\xi/2} \end{array}\right), \qquad
\tl X^{A}{}_B = - \f{(\g+i)\veps j}4 \left(\begin{array}{cc} -1 & 0 \\ 0 & 1 \end{array}\right).
\ee
We notice that the translations are removed dividing by the $F_2$ orbits. The same happens in the representation of massless particles, and here it has the nice geometric interpretation of being shifts along a null direction.
The remaining algebra consists of the helicity generator $L^3$, which coincides with the oriented area of the bivector,
\be\label{Lj}
L^3 = \veps j = -\tl L^3, \qquad \{ L^3, \xi \} = 1 = -\{\tl L^3, \xi\}.
\ee
We conclude that $\T_2/\!/C/\!/F=T^*\SO(2)$, parametrized by its holonomies and fluxes, or directly by $(\veps j,\xi)$.
After symplectic reduction, the initial Lorentz algebra has collapsed to the helicity subgroup $\SO(2)$ of $N^I$. In particular, $\veps$ is the sign of the helicity, consistent with its initial twistorial definition, \Ref{gij}.

Let us also discuss the covariance of our construction. Above we have fixed the same null vector 
 for both source and target nodes, $N^I = \tl N^I = (1,0,0,1)/\sqrt{2}$, and the reduction has led to the canonical little group. Any different choice, say for the source, can be written as $V N$, where $V$ is a group element in the complement of the little group, and similarly $\tl V\tl N$ for the target normal. In this general case, the resulting reduced phase space would be of the form 
 $(V X V^{-1}, V g \tl V^{-1})$, 
 that is the canonical little group embedded by the conjugate action.
In this sense, our construction is completely covariant.

\section{Null twisted geometries}

We have so far described the constraint structure and the symplectic reduction on a given link. 
We now move on to consider the full graph, and include the closure condition \Ref{Gn} in the analysis.
For simplicity, we take the same canonical null vector $N^I$ on each node. The case of arbitrary $N^I$ can be dealt with via the adjoint action as explained above, and does not change the geometric interpretation which is covariant by construction.
The results of the previous section show that the twistor phase space on the graph, reduced by the null simplicity conditions \Ref{linsimpl} and the area matching \Ref{defC}, is
$\T^{2L}/\!/C_l/\!/F_{nl} = T^*\SO(2)^L$, a phase space of dimensions $2L$, parametrized by $(\veps_l j_l, \xi_l)$. 
This result used the fact that the simplicity constraints are all first class by themselves.
The situation slightly changes when the closure condition\Ref{Gn} is included. On shell of the simplicity and area-matching constraints, \Ref{Gn} reduces to
\be\label{Gnred}
G_n = \sum_{l\in n} L^3 =0, \qquad \wh I^a_n = \sum_{l\in n} \wh P^a=0, \qquad a=1,2.
\ee
Here $\wh P^a$ are the translation generators of the little group of $\hN^I=\mathcal{P}N^I$, the only generators changing $N^I$.

These three conditions are equivalent to \Ref{eq:CloseC}, in particular the first is the area closure \Ref{areaClos}, as follows immediately from \Ref{b} and \Ref{Lj}. Taking into account the link orientations, we have
\be\label{U1clos}
G_n = \sum_{l^+\in n} L^3 + \sum_{l^-\in n} \tl L^3
= \sum_{l^+\in n} \veps_l j_l - \sum_{l^-\in n} \veps_l j_l =0,
\ee
where $l^+$ are the links outgoing from the node, 
and $l^-$ the incoming ones.
This expression coincides with the area closure \Ref{areaClos}, once we take into account that $\veps_l$ coincides with the time orientation for an outgoing link, and its opposite for an incoming link, as discussed below \Ref{tpisol}.
Therefore, we can interpret the reduced phase space as a collection of null polyhedra, dual to the nodes of the graph. The polyhedra are glued along faces, sharing the same area $A_l \propto j_l$, and with opposite time orientation.

Notice that out of the closure conditions \Ref{Gnred}, only $G_n$ generates an isometry of the null plane. The other isometries of the null hypersurface are not generated by the closure condition, but by combinations of the simplicity constraints, as can be deduced from their action investigated in the previous section, and to which we will come back below.
As it turns out, $\wh I^a$ do not generate symmetries at all, as they form a second class system with part of the $F_2$ simplicity constraints.\footnote{Notice that in the timelike case, the covariant closure condition is a first class constraint in the discrete theory, whereas the continuous Gauss law in the time gauge has a second class part corresponding to the complement to the little group. In this sense, the null case considered here bears some interesting similarities with  the continuum theory.}
To study the structure of the constraints and bring this fact to the surface, we compute the Dirac matrix associated with the graph. 
As variables on different links commute, the matrix has a block structure, in which each block is associated with a node. 
Since the Lorentz-invariant constraints $F_{1}$ commute with everything, we leave them out of the analysis. 
Then for a node of valence $m$, the $F_2$ and closure constraints form a $(2m+3)$-dimensional system. 
On shell of the $F_1$ constraints, it is possible and convenient to replace for each link the complex $F_{2}$ constraints by the two real $P^a$. We then take the basis of node constraints
\be\label{defphi}
\phi_\mu = \{ P_1^1, P_2^2, \ \ldots \  , P_m^1, P_m^2, \wh I^1, \wh I^2, G \}.
\ee
On the constraint surface, the node's block of the Dirac matrix evaluates to 
\begin{equation}
D_{\mu\nu}\equiv\{\phi_\mu, \phi_\nu\}  \approx 
\left(\begin{array}{ccccc|ccc}
0 & 0 & \cdots & 0 & 0 & -2\gamma L_1^3 & 2 L_1^3 & 0 \\
0 & 0 & \cdots & 0 & 0 & -2 L_1^3 & -2\gamma  L_1^3 & 0 \\
\vdots & \vdots & \ddots & \vdots & \vdots &\vdots & \vdots & \vdots \\
0 & 0 & \cdots & 0 & 0 & -2\gamma L_m^3 & 2 L_m^3 & 0 \\
0 & 0 & \cdots & 0 & 0 & -2 L_m^3 & -2\gamma  L_m^3 & 0 \\
\hline
2\gamma L_1^3 & 2 L_1^3 & \cdots & 2\gamma L_m^3 & 2 L_m^3 & 0 & 0 & 0\\
-2 L_1^3 & 2 \gamma L_1^3 & \cdots & -2 L_m^3 & 2 \gamma L_m^3 & 0 & 0 & 0\\
0 & 0 & \cdots & 0 & 0 & 0 & 0 & 0\end{array}\right)
\end{equation}
The rank of this matrix is always 4, independent of the valence of the node. Hence, the node algebra contains 
$2m-1$ first class constraints and two pairs of second class constraints. 
Using this result, and reintroducing the $F_1$'s (one independent first class constraint per link), the counting of dimensions of the reduced phase space $\cS_\G$ defined in \Ref{SG} gives
\be
12L-2L-4N-2\sum_n(2 \, {\rm valence}_n-1) = 2L -2N.
\ee
It is much smaller than in the timelike case, where one obtains $6L-6N$, which we recall to the reader that it represents a collection of Euclidean polyhedra plus an angle ($\xi$ in the literature) associated with each shared face. In the null case, the reduced space is much smaller. Since we proved at the beginning of the paper that a geometric interpretation in terms of null polyhedra is still possible, we must conclude that information on the intrinsic shapes of the polyhedra is being lost in the reduction. In fact, recall from \Ref{Totti} that on each face the orbit of $F_2$ changes the value of $b^I$. These transformations can be distinguished in three types. First, those corresponding to translations of the vertices in the null direction, which correspond to isometries. Second, those corresponding to translations of the vertices changing the reconstructed angles \Ref{angles}, and thus the intrinsic geometry of the polyhedron. Third, those incompatible with the closure condition \Ref{Gnred} and thus breaking the polyhedron apart.
The first two types turn out to be first class, while the third type is second class. Therefore, while the interpretation in terms of closed polyhedra is valid, because of the closure condition, the intrinsic shapes at fixed areas are pure gauge, the variables $\om^0_l$ drop out, and the reduced phase space contains only the conjugated variables $(\eps_l j_l,\xi_l)$, constrained by the first class constraint $G_n$. Hence,
\be
\cS_\G = T^*\SO(2)^L/\!/G_n.
\ee
We now prove these statements.

To diagonalize the Dirac matrix on each node, we first observe that the combinations
\begin{align}\label{F2First}
& C^a_{ij} := L^3_i P^{a}_j-L^3_j P^{a}_i=0, \\
& I^a := \sum_{l\in n} P^a=0 \label{GnF2First}
\end{align}
are first class. Second, the set
\be
C^a_{1i}, \quad i=2,3,\cdots,m-1, \qquad P^a_m, \qquad I^a
\ee
is equivalent to all of the $F_2$'s. Therefore, we can take out of \Ref{defphi} the two pairs $(P^a_m, \wh I^a)$ as the four second class constraints, and the rest are first class, with $P_1^a,\ldots ,P_{m-1}^a$ replaced by (\ref{F2First}) and (\ref{GnF2First}). 
In particular, the first class constraints contain the global isometry $\ISO(2)$ generated by $I^a$ and $G_n$,\footnote{The remaining isometry of the null hypersurface, the boosts $\sum_l K^3_l$, is generated by the $F_1$'s.} 
as well as $2m-4$ additional first class constraints. Their orbits can be used, together with the four second class constraints, to eliminate all of the $\om^0_l$ from the reduced phase space.

To see this explicitly, we compute the action of the first class generators on the spinors, obtaining
\begin{equation}
\ex^{\{-\alpha_{j} (\ii C^1_{1j}-C^2_{1j}),\bullet\}}\omega_i^0=\omega_i^0+ \d_{ij} \l_j \omega_i^1, 
\qquad \l_i := \alpha_i(\gamma+\ii)\varepsilon_i j_i, \qquad i=2,\cdots, m-1,
\ee
and
\be
\qquad \ex^{\{-\beta (\ii I^1-I^2),\bullet\}}\omega_i^0= \omega_i^0+\beta \om_i^1. 
\end{equation}
Therefore, we can always set to zero all $\om^0_l$, except when $l=m$. The remaining variable is, however, constrained by the second class closure constraint in \Ref{Gnred},
\begin{equation}
\omega^0_m=- \frac{z_m |\omega^1_m|^{\ii\gamma+1}}{\varepsilon_m j_m^{3/2}}~~\sum_{i=1}^{m-1}\varepsilon_ij_i^{3/2}\frac{\omega_i^0}{z_i|\omega_i^1|^{\ii\gamma+1}},
\end{equation}
and it is thus automatically vanishing with the previous gauge choice.

Going back to the picture of the null tetrahedron, we see that there are some constraints which generate the global isometries, and others which can arbitrarily move around the vertices of the polyhedron, while preserving the closure and the individual areas. In doing so, we can squash the polyhedron on the spacelike surface and wash away as gauge all information on the intrinsic shapes. 
This becomes manifest if we rewrite the null polyhedra in terms of the reduced variables. 
To see this, we fix the $F_1$ gauge $|\omega^1|=1$ and write the spinors in terms of $z_l$ and the orbits of $C^a_{1i}$ and $I^a$, 
\begin{equation}
\omega_i^A=\left((\lambda_i+ \beta)\ex^{\ii\arg(z_i)}, \ex^{\ii\arg(z_i)}\right), \qquad i\neq 1, m,
\end{equation}
and the $\pi_i^A$ are given by \Ref{pi=rom}, assuming all the links are outgoing. Let us consider the case of a 4-valent node, so we do not have to deal with the reconstruction procedure, and we can immediately apply the formulas \Ref{lengths}. 
A straightforward calculation then gives
\begin{equation}
 E_{12}^2=\gamma\frac{j_1j_2}{3j_3}\frac{|2\lambda_2+\lambda_3|^2}{\im(\lambda_2\bar{\lambda}_3)}, \qquad
 E_{12} \cdot E_{23} = -2\gamma\varepsilon_1\varepsilon_3j_2 \frac{|\lambda_2|^2+|\lambda_3|^2+\re{\l_2\bar{\l}_3}}{\im(\lambda_2\bar{\lambda}_3)}.
\end{equation}
The intrinsic shape of the null tetrahedron is determined by the independent areas and also the gauge orbits of $C_{1i}$, while being invariant under action of the isometries, in particular $\beta$ drops out.

\subsection{Intrinsic geometry: Euclidean singular structures}
We have seen above that the first-class constraints eliminate the intrinsic shapes at fixed areas and we are left with an Abelian reduced phase space $T^*\SO(2)$.
The remaining closure condition \Ref{U1clos} can be solved explicitly, and we are able to provide a complete set of gauge-invariant observables, unlike in the non-Abelian case. This leads to a very simple geometric picture, where the polyhedra give way to a continuous, albeit singular, metric structure. 

Consider a closed graph, the extension to an open graph being straightforward. 
The dimension of the reduced phase space is $2(L-N+1)$, where we took into account the fact that on a closed graph one of the closure conditions is redundant.
The gauge invariant information can be associated with the faces of the graph, up to moduli taking into account the possible nonplanarity of the graph.
Consider first a planar graph. Its genus being zero, $2(L-N+1)=2(F-1)$,
so it is enough to remove the pair of variables associated with a specified face, say for instance the external one in the Schlegel representation of the graph.
Denoting $f=1, \ldots F-1$, we trade the $\xi_l$ for the gauge-invariant traces of the holonomies, 
\begin{subequations}\label{gicoord}
\be
\Phi_f := 2 \arccos \bigg[\f12 \tr \bigg( \prod_{l\in\p f} h_l \bigg)\bigg] \approx \sum_{l\in\p f}\eta_l \xi_l, \qquad 
\{G_n, \Phi_f \}=0,
\ee
where $\eta_l=\pm$ depending on the consistency of the orientation between the face and the link.
The same faces can be used to define an independent set of spins, 
\be
J_f := \sum_{l\in\p f} \eta_l j_l.
\ee
\end{subequations}

The reason to weigh the sum with the same signs is to have a nice Poisson structure. In fact, for a planar graph the faces can be consistently oriented so that each link is traversed in opposite directions by the sharing faces. A moment of reflection then reveals that the coordinates \Ref{gicoord} of the gauge-invariant phase space satisfy the brackets
\be\label{JPPB}
\{J_f, \Phi_{f'} \} = L_{ff'},
\ee
where $L_{ff'}$ is the Laplacian of the dual graph.\footnote{Notice that this graph is open, because of the redundancy of a global closure condition and associated gauge.}
Proper action-angle variables can then be readily found diagonalizing the Laplacian.

\begin{figure}[!htbp]
\centering
\includegraphics[width=\textwidth]{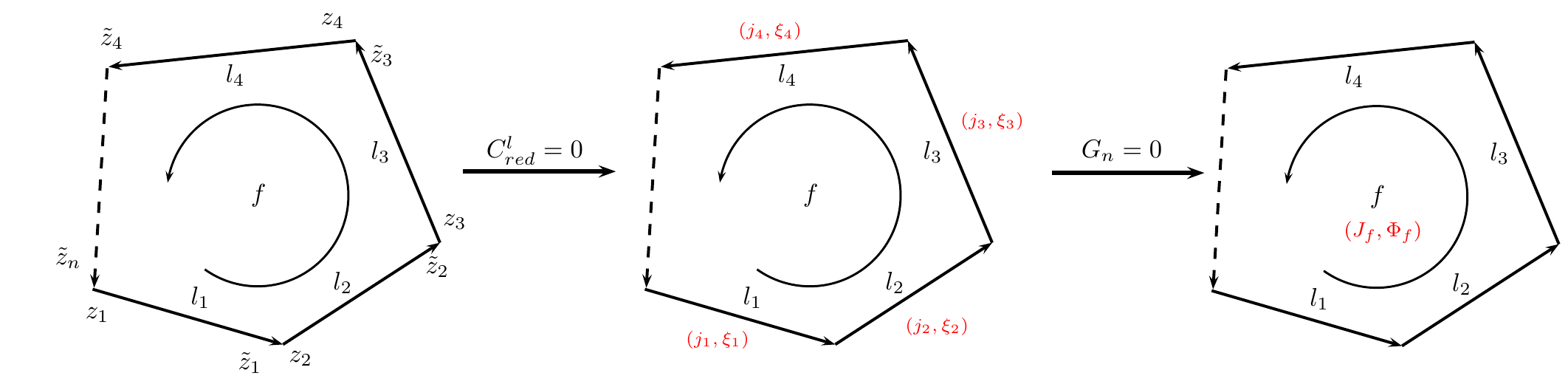}
\caption{From half links $(z,\widetilde{z})$ to links $(j,\xi)$ and to loops $(J,\Phi)$}
\label{zjJ}
\end{figure}

Since the intrinsic shapes of the polyhedra have been gauged away, the reduced variables describe equivalence classes characterized uniquely by the areas. However, the same variables can be given a simpler and more direct geometric interpretation. Recall that the intrinsic geometry is fully determined by the projection on $S_0$. One can then describe a spacelike 2d geometry using the reduced variables. First of all, we observe that the reduced gauge-invariant holonomies describe an $\SO(2)$ transformation on each face. For simplicity, consider first the case of a trivalent graph dual to a triangulation. This structure alone defines the conformal structure of a 2d Regge geometry, that is a collection of deficit angles $2\pi-\Phi_f$ associated with the vertices dual to the faces. Then, the positive real number $J_f$ associates a scale with each face, thus picking a representative of the conformal class. 
\begin{figure}[!htbp]
\centering
\includegraphics[width=0.8\textwidth]{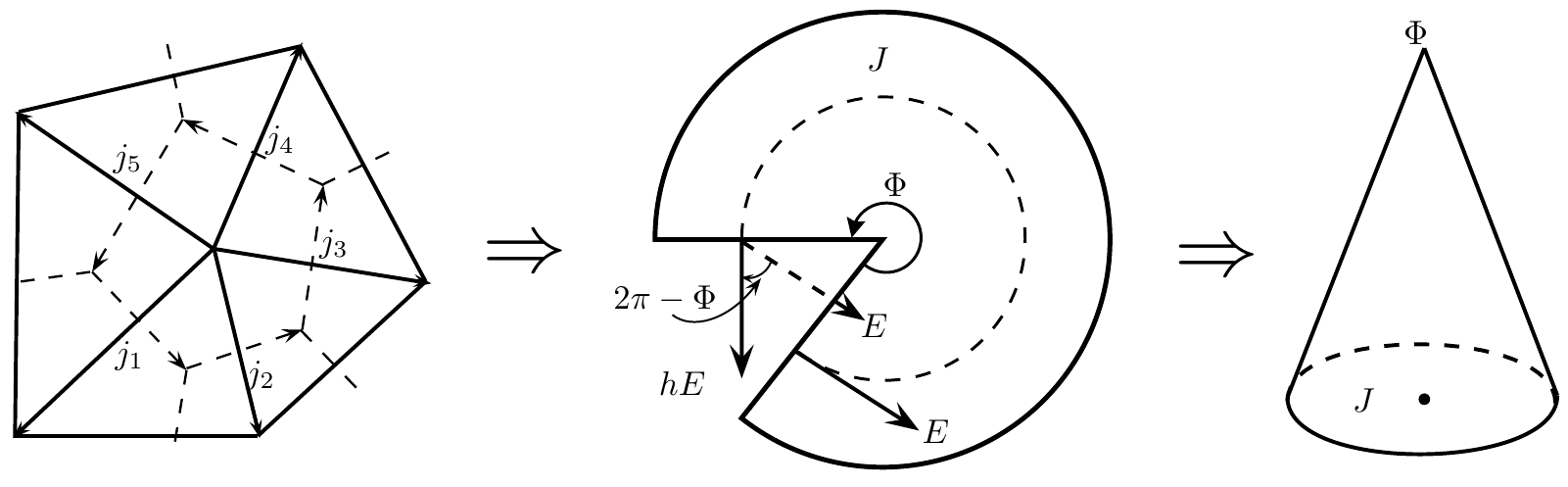}
\caption{The deficit angle $(2\pi-\Phi)$ and the scale $J$ of the cone}
\label{djcone}
\end{figure}
If we pick a local complex coordinate on each face, say $\z_f$, chosen so that the origin is the location of the vertex, we can write the face metric as
\be
ds^2 = J_f \, |\z_f|^{-{\Phi_f}/{\pi}} \, d\z\otimes d\bar\z.
\ee
The resulting geometry is a singular Euclidean structure 
(e.g. \cite{Carfora:2002rn}) on $S_0$. 

Notice that by assigning these variables we are specifying fewer data than those required by a 2d Regge triangulation, which would be $L=3(F-2)$. A Regge geometry would be specified uniquely if instead of assigning a scale factor to each dual face, we would do so to each triangle. Since a triangulation has more triangles than vertices, our data are fewer and do not specify a unique 2d Regge geometry. 
On the other hand, it is more general than a Regge geometry in the sense that it can be extended to any graph and not just a dual to a triangulation, and furthermore because the special case $\Phi_f=2\pi$, which in Regge would be a pathological infinite spike, is a perfectly regular configuration, which can be interpreted as hyperbolic triangles \cite{Carfora:2002rn}.
Finally, the description has the pleasant features of a natural split into a conformal metric plus scale factors, locally conjugated.

For non-planar graphs, the situation is slightly different, because more than the faces, one should look at the independent cycles, and these cannot be oriented in such a way that each link is traversed at most twice, in opposite directions. 
Therefore evaluation of Poisson brackets gives a matrix whose off-diagonal entries can have both signs. This can \emph{a priori} still be interpreted as a weighted Laplacian of some dual graph, but one in which the weights have indefinite signature. For instance, in the case of the 4-simplex, the six independent cycles can be chosen so that there is a single $-1$ entry in the adjacency matrix.\footnote{The cycles are e.g. $012, 103, 132, 402, 430, 413$, and the Poisson brackets evaluate to the following matrix, 
\be
\left(\begin{array}{cccccc}
3 & -1 & -1 & -1 & 0 & 0 \\
-1& 3 & -1 & 0 & -1 & -1 \\
-1 & -1 & 3 & 0 & 0 & 1 \\
-1 & 0 & 0 & 3 & -1 & 0 \\
0 & -1 & 0 & -1 & 3 & -1 \\
0 & -1 & 1 & 0 & -1 & 3
\end{array}\right).
\ee
It can still be casted in the form $D-A$ of a certain dual graph, where $D$ and $A$ are respectively the degree and weighted adjacency matrix, with the latter having also negative entries.}

\subsection{Extrinsic geometry: $\Xi$ and the role of the embedding}
The above description concerns the intrinsic geometry of the hypersurface, which being null is equivalent to a 2d one. 
However the 3d nature should show up in the study of the extrinsic geometry. As the reader familiar with loop quantum gravity knows, information on the extrinsic geometries is also contained in the reduced phase space, but it is mixed with the intrinsic one. This is the trade-off for the use of real Ashtekar-Barbero variables. It can be extracted once the solution to the secondary simplicity constraint is known, for this provides a specific (in general, nontivial) embedding of the reduced phase space into the Lorentzian one. The same has been argued to happen in the discrete theory in \cite{IoWolfgang}, and indeed shown 
at least for flat dynamics \cite{IoFabio}. A similar situation should happen in the present null case, and in order to talk about extrinsic geometry, we need to first understand the dynamics of our null twisted geometries, which we plan to do in future work. 

Here we limit ourselves to characterizing the kinematical degrees of freedom suitable to describing the extrinsic geometry.
In the timelike case, this was identified on the constraint surface as the (boost) dihedral angle between the normals $N^I$ in adjacent nodes. However, as we stressed above in \Ref{hred}, in the null case the holonomy is a restricted group element already at the level of the constraints surface, and as a consequence, the angle between the normals $N^I$ and $\tl N^I$ on adjacent nodes vanishes,
\be
\tl N \cdot \L(h) N = 0.
\ee
The vanishing of this scalar product is consistent with the fact that we are dealing with a null hypersurface, and in order to specify a notion of extrinsic geometry, we need an embedding in some nondegenerate four-dimensional spacetime. 
Indeed, considering also the null hypersurface spanned by the parity transformed vector $\wh N^I$, we can evaluate a nonzero scalar product, given by
\be
\mathcal{P}\tl N\cdot \L(h) N = -\ex^{\Xi},
\ee
where $\Xi$ is the boost rapidity previously defined, and $\L(h) N = \ex^{\Xi} N$.
The equation above suggests that $\Xi$ should be related to a discretization of a certain free coordinate (denoted $\l$ in \cite{MikeNull07}) used in the null formulation of general relativity \cite{Sachs62,MikeNull07,MikeNullPRL}. We postpone the comparison of our discrete data to a discretization thereof to future work.

We expect that $\Xi$ plays an important role in characterizing the extrinsic geometry, as well as possibly the intrinsic shapes of the null polyhedra. The fact that these quantities have disappeared from the reduced phase space has do to with the fact that in the constrained system considered so far, the simplicity constraints were all first class. Future studies of the dynamics may reveal the presence of secondary constraints, that could turn some or all of the simplicity constraints into second class, e.g. \cite{alexandrov2008simplicity}. If that happens, the solutions to the secondary constraints can be interpreted as providing specific, nontivial gauge fixing for the orbits, thus restoring a geometric interpretation for $\Xi$ and the intrinsic shapes through the dynamical embedding.

\section{Quantization and null spin networks}

Quantizing the above phase space and its Poisson algebra introduces a notion of spin networks for null hypersurfaces. 
The reduced phase space $T^*\SO(2)$ with its canonical algebra $\{m,\xi\}=1$, $m=\veps j$, can immediately be quantized on the Hilbert space $L_2[\SO(2)]$, the space of SO(2) unitary irreducible representations with eigenvalues $m\in\Z/2$, and operator algebra
\be\label{hfSO2}
\psi[\xi], \qquad [\hat m, e^{\ii\hat\xi/2}] = \f12 e^{\ii\hat\xi/2}.
\ee
Since $\xi\in[0,4\pi)$, the eigenvalues of $\hat m$ are half-integers, and $e^{\ii\hat\xi}$ acts as a raising operator, 
\be
\hat m \ket{m} = m \ket{m}, \qquad e^{\ii\hat\xi/2}\ket{m} = \ket{m+1/2},
\ee
the Abelian version of the holonomy-flux algebra.
Finally, a basis is given by Fourier modes on the (double cover of the) circle,
\be
\psi_m[\xi] = \bra{\xi} m\ra = e^{\ii m \xi}.
\ee
This Hilbert space bears similarities with the more familiar one of the harmonic oscillator in action-angle variables, the main difference being that the ``Hamiltonian" $\hat m$ is not bounded from below, and $m\in\Z/2$.

The gauge-invariant Hilbert space $\hh_\G$, corresponding to ${\cal S}_\G$, is obtained by taking the tensor product of the states on the links and imposing the closure condition \Ref{U1clos} on the nodes. 
The results are Abelian $\SO(2)$ spin networks, with trivial intertwiners and flux conservation on the nodes,
\be\label{u1spinnets}
\Psi_{\G,m_l}[\xi_l] = \otimes_l \psi_{m_l}[\xi_l] \prod_n \d\Big(\sum_{l^+\in n} m_l - \sum_{l^-\in n}  m_l \Big).
\ee

To appreciate how these simple states can represent quantized null hypersurfaces, it is instructive to derive $\hh_\G$ following Dirac's procedure, starting from a Hilbert space for the twistor phase space and its algebra, and then implement the quantized constraints.
This procedure will show how such Abelian spin networks are to be embedded in the Lorentz group, and identify $m$ as the helicity quantum number.
While being necessary for future studies of dynamics, it will also expose some of the covariance properties of the states, as well as their integrability properties with respect to the $\SL(2,\C)$ Haar measure.
As in the classical reduction, we proceed in two steps: we first consider the quantization of a single twistor phase space, and the simplicity constraints it satisfies; then, we look at the link phase space and impose the area-matching condition.

For the twistorial Hilbert space we take wave functions $f(\omega)\in L^2[\mathbb{C}^2,\dd^4\omega]$, where
\begin{equation}\label{d4om}
\dd^4\omega=\f{1}{16}\dd\omega_A\wedge\dd\omega^A\wedge cc,
\end{equation}
and a Schr\"{o}dinger representation of the canonical Poisson algebra (\ref{eq:CPoiB}),
\begin{equation}
 [\hat\pi_A,\hat\omega^B]=-\ii\hbar\delta_A^B, 
\qquad (\hat\omega^A f)(\omega^A)=\omega^A f(\omega^A), \qquad (\hat\pi_A f)(\omega^A)=-\ii\hbar\frac{\partial}{\partial\omega^A}f(\omega^A).
\end{equation}
A convenient basis for these is provided by homogeneous functions, since they diagonalize the dilatation operator appearing in $F_1$, and carry a unitary, infinite-dimensional representation of the Lorentz group. In particular, since the simplicity constraints are the vanishing of the $\ISO(2)$ translation generators $P^a$, it is convenient to take a basis diagonalizing the latter, called the null basis, instead of the canonical basis labeled by the rotational subgroup $\SU(2)$. Denoting $p^a$ the eigenvalues, and $p:=-p^2+\ii p^1$, the null basis element are the wave functions
\begin{equation}
f_p^{(\rho,k)}(\omega^A)=\frac{1}{2\pi}(\omega^1)^{-k-1+\ii\rho}(\bar{\omega}^{\dot{1}})^{k-1+\ii\rho}\exp\left[\frac{\ii}{2}\left(\frac{\bar{\omega}^{\dot{0}}}{\bar{\omega}^{\dot{1}}}p+\frac{\omega^0}{\omega^1}\bar{p}\right)\right]
\end{equation} 
where $(\rho\in\mathbb{R},k\in\mathbb{Z}/2)$.
Details about the $\sltc$ and ISO(2) representations can be found in the Appendix. 

To represent quadratic operators, we introduce the normal ordering
\begin{equation}\label{eq:NormalOrder}
:\wh\po:  = \frac{1}{2}(\hat\pi_A\hat\omega^A+\hat\omega^A\hat\pi_A) =-\ii\hbar\left(\omega^A\frac{\partial}{\partial\omega^A}+1\right).
\end{equation} 
With this ordering, the spinorial simplicity constraints \Ref{Fdef} read
\begin{equation}
\hat F_1=\frac{\hbar}{2}\left((\gamma-\ii)\omega^{A} \frac{\partial}{\partial\omega^{A}}-(\gamma+\ii)\bar{\omega}^{\dot{A}} \frac{\partial}{\partial\bar{\omega}^{\dot{A}}}-2\ii\right), \qquad
\hat{F}_2=\ii\hbar\bar{\omega}^1\frac{\partial}{\partial\omega^0}, \qquad 
\hat{\bar{F}}_2=\hat F_2^\dagger = \ii\hbar\omega^1\frac{\partial}{\partial\bar\omega^{0}}.
\end{equation}
Since on each link these constraints are first class, they can be imposed as operator equations on states. An immediate calculation then gives
\begin{align}
& \hat{F}_1 f_{p}^{(\rho,k)}(\omega_A) = 0  \quad \Rightarrow \quad \rho=\g k, \\
& \hat{F}_2 f_{p}^{(\rho,k)}(\omega_A) = \hat{\bar F}_2 f_{p}^{(\rho,k)}(\omega_A) = 0  \quad \Rightarrow \quad p=0,
\end{align}
so the solutions are the functions 
\begin{equation}
f_k(\omega^A)\equiv f^{(\gamma k,k)}_0(\omega^A)=\frac{1}{2\pi}(\omega^1)^{(\ii\gamma-1)k-1}(\bar{\omega}^{1})^{(\ii\gamma+1)k-1}.\label{sols}
\end{equation}
The formula \Ref{sols} defines a state also for $k=0$, but this case corresponds classically to $\po=0$, for which the twistorial description of $T^*\SL(2,\C)$ breaks down. To complete the quantization, we need to provide independently the missing state. If we extrapolate \Ref{sols} to $k= 0$ we get a nontivial state, $|\omega^1|^{-2}$, which could pose problems with cylindrical consistency. Hence, we fix instead
\begin{equation}
f_0(\omega^A)=1.
\end{equation}

The first thing to notice is that in the $p=0$ sector $P^a$ and $L^3$ commute, thus these functions are also eigenfunctions of $L^3$, with
\begin{equation}
\hat{L}^3 f_k(\omega^A)=\hbar k f_k(\omega^A),
\end{equation}
and thus $k$ is the helicity eigenvalue.
Next, the solutions can be expressed in terms of the reduced phase space variable $z$ using \Ref{defz}, obtaining 
\be\label{fkom}
f_k(\om^A) = \f1{2\pi |\om^1|^2} \left( \f{\bar z}z\right)^k.
\ee
Notice the leftover dependence on the non-$F_1$-invariant term $|\omega^1|$. As the action generated by $F_1$ is noncompact, Dirac's quantization does not lead to a proper subspace of functions on the reduced phase space, but rather distributions. Proper function can be defined taking into account the reduced measure.

The reduced measure can be obtained starting from \Ref{d4om}, imposing the constraints and dividing by the gauge orbits generated by their Hamiltonian vector fields $h_{F_i}$,
\be
\dd\mu(z) := 4\pi\ii \, \iota_{h_{F_i}}(\dd^4\om){\big|_{F_i=0}},
\ee
where $\iota$ denotes the interior product and $4\pi\ii$ is a normalization motivated \emph{a posteriori}.
The Hamiltonian vector fields are
\be
h_{F_1} := \{F_1, \bullet \} \approx \f12(1+i\g)\om^0 \f{\p}{\p\om^0} + \ii \g  \om^1 \f{\p}{\p\om^1} +cc.
\qquad h_{F_2} := \{F_2, \bullet \} \approx -2 \om^1 \f{\p}{\p\om^0}.
\ee
Evaluating the interior products gives
\be
\iota_{h_{F_2}} \iota_{h_{\bar F_2}} [(\dd\omega_A\wedge\dd\omega^A)\wedge cc.] 
\approx - 4 {|\om^1|^2}\,  \dd\om^1\w \dd\bar\om^1,
\ee
and 
\be
\iota_{h_{F_1}} (\dd\om^1\w \dd\bar\om^1) \approx \ii\g (\om^1\dd\bar\om^1-\bar\om^1\dd\om^1).
\ee
Putting these results together, and reintroducing $z$, we get
\be
\dd\mu(z) = -\pi{\ii} |\om^1|^4\, \left(  \f{\dd\bar z}{\bar z} - \f{\dd z}z \right).
\ee
Notice that the dependence on $\g$ has disappeared, and the measure factor $|\om^1|^4$ perfectly compensates the one in the reduced functions \Ref{fkom}. 

Denoting $\arg(z)=-2\phi$, we have 
$\dd\mu(z)=4 \pi |\om^1|^4 \dd\phi$, and the proper reduced Hilbert space is given by
\be
f_k(\phi) = \bra{\phi}k\ra = \f1{2\pi}  e^{2\ii k\phi},
\qquad \bra{k'}k\ra = \f1{\pi} \int_0^{\pi} \dd \phi \, e^{2\ii (k-k')\phi} = \d_{kk'},
\ee 
with $k\in\Z/2$. 
This half-link Hilbert space already coincides with $L_2[\SO(2)]$,  with operator algebra 
\be
\hat m \ket{k} = k\ket{k}, \qquad \exp\left(\ii\frac{\hat{\phi}}{2}\right) \ket{k} = \ket{k+\f12}.
\ee 

The next step is to consider the two copies of this Hilbert space associated with a link, and impose the area-matching condition,
but this procedure will lead trivially to an equivalent Hilbert space.\footnote{This should not come as a surprise: the whole point of the twistorial parametrization is to encode a nonlinear space (the group manifold) into the solution to a quadratic equation of a linear space (twistor space). But if the starting point is already linear, as in this Abelian case, the procedure is clearly trivial.}
In fact, the quantum version of the area-matching condition on one link corresponding to (\ref{defC}) is
\begin{equation}
\hat{C}\equiv:\wh\po:+:\wh\tpo:
\end{equation}
and imposing it strongly on a tensor product state $f_k(\omega^A)\otimes f_{\widetilde{k}}(\to^A)$ gives immediately $k=-\widetilde{k}$. The state simplifies to
\be\label{Fxi}
F_k(\xi) = \f1{(2\pi)^2} e^{\ii k \xi}, \qquad \xi\in[0,4\pi).
\ee
The appropriate link measure is also obtained trivially.
We have thus recovered the initial $L_2[\SO(2)]$, with holonomy-flux algebra \Ref{hfSO2}, and further we can identify the oriented area operator $\hat m$ with the helicity and its eigenvalues with the label $k$ of the Lorentz irreps. 

Finally, gauge invariance can easily be implemented, and the results are the Abelian spin networks \Ref{u1spinnets}.
Just as ordinary $\SU(2)$ spin networks can be interpreted as quantized twisted geometries, the null spin networks represent quantized null twisted geometries.\footnote{In other words, coherent states of \Ref{u1spinnets} are peaked on a null twisted geometry. }

The embedding allows us to define and evaluate generic Lorentz operators on the reduced Hilbert space. For instance, the first Casimir, classically the oriented area 
\be
A^2 = \frac{1}{2}B_{IJ}B^{IJ}=
\frac{\g^2}{2(\gamma^2+1)^2}\left[(\gamma-\ii)^2(\po)^2 +(\gamma+\ii)^2(\bar\po)^2\right]
\approx \g^2 j^2,
\ee
is the last equality holding onto the constraint surface. 
The corresponding operator is
\begin{equation}
\hat{A^2}\equiv\frac{-\gamma^2\hbar^2}{2(\gamma^2+1)^2}\left[(\gamma-\ii)^2\left(\omega^A\frac{\partial}{\partial\omega^A}+1\right)^2 +(\gamma+\ii)^2\left(\bar\omega^{\dot A}\frac{\partial}{\partial\bar\omega^{\dot A}}+1\right)^2\right], 
\end{equation}
and on the solution space spanned by \Ref{Fxi} gives
\begin{equation}
\hat{A^2} F_k = \hbar^2 \g^2 k^2 F_k.
\end{equation}

\section{Discussion}
In this paper, we have exploited the parametrization of LQG on a fixed graph in terms of twistors to describe null hypersurfaces and their quantization in terms of spin networks. Our construction is based on the fact that the twistors appearing in LQG satisfy a restricted incidence relation, in turn determined by the timelike vector appearing in the $3+1$ decomposition of the Plebanski action. Taking this vector to be null forces the geometric interpretation of the theory to lie on a null hypersurface, and the result is a collection of null polyhedra with spacelike faces. 

The first result of our paper concerns properties of the geometry of null polyhedra. We provided a characterization of the intrinsic shapes in terms of simple bivectors, and showed that the space of shapes at fixed external areas is not a phase space obtained from bivectors and the action generated by the closure constraint, as it is the case for spacelike and timelike polyhedra, because in the null case the reduced closure condition does not generate all of the isometries, but only the helicity part of it. The rest of the closure is second class. The remaining isometries are in turn generated by the (global) action of the simplicity constraints around a node. However, all the simplicity constraints (compatible with the closure condition) are first class, not just their total sum on a node, and their action changes the intrinsic shapes of the null polyhedron. Therefore, the phase space obtained by symplectic reduction is much smaller, algebraically described just by the helicity subgroup, and geometrically an equivalence class of null polyhedra determined only by the areas and their time orientation.

The second result concerns the description of the gauge-invariant phase space. As the helicity subgroup is Abelian, the remaining closure condition can be solved explicitly, and proper action-angle variables given. For planar graphs, these are given by the eigenvectors of the Laplacian of the dual graph. The action-angle variables have a compelling geometric interpretation, as a Euclidean singular structure on the two-dimensional spacelike surface determined by a null foliation of spacetime. In particular, it is naturally decomposed into deficit angles and scale factors, locally conjugated.
We are not in a condition to discuss the extrinsic geometry and thus the three-dimensional picture of the null twisted geometries, because this requires the discrete analogue of the secondary simplicity constraints, and it is thus referred to future work on the dynamics. However, we identified the variables in the phase space susceptible of carrying such information.

Finally, we quantized the phase space and its algebra, introducing a notion of null spin networks. They are Abelian spin networks, whose embedding the Lorentz group permits one to identify the Abelian quantum number with the helicity along the null direction of the hypersurface. We derived the spin networks by directly quantizing the reduced phase space, and also by following Dirac's procedure starting from a Hilbert space for twistors.
Notice that a loop-inspired quantization of null hypersurfaces has appeared some time ago in \cite{CarloNull96}. The main difference is that the approach of \cite{CarloNull96} is based on asymptotic quantities defined at null infinity, whereas here we look at local quantities associated with a fixed graph. Notwithstanding this important difference, a comparison of the two approaches would be valuable.

As such, our result are only a first, kinematical step toward our goal of understanding the dynamics of null surfaces in LQG. 
The applications are many and furnish important motivations to our research program,  from the possibility of including dynamical effects in black hole physics and isolated horizons \cite{AshtekarQIH}, describing the near horizon quantum geometry, to the use in the constraint-free formulation of general relativity on null hypersurfaces.
To that end, many nontivial steps are needed. 
First of all, our analysis needs to be complemented with a continuum canonical analysis of the Plebanski action on a null hypersurface \cite{IoAlexandrov}.
Second, our geometric description should be compared with the null formulations of general relativity  \cite{MikeNull07,Sachs62,MikeNullPRL,FrittelliNull95}, and suitable discretizations thereof,
in particular, identifying the shear degrees of freedom, and completing the geometric picture developed here with its extrinsic geometry. 
On a complementary level, one should also investigate what type of spin foams can support the boundary data here studied (see e.g. \cite{Neiman:2012fu}).
We expect this line of research to bring new tools and results to LQG, and to show us how deep the connection with twistors goes.

\subsection*{Acknowledgements}

We are grateful to Sergey Alexandrov, Hal Haggard, Mike Reisenberger, Carlo Rovelli and especially Wolfgang Wieland for many discussions and valuables comments.
Simone would like to thank the organizers and participants of Peyresq Physics 18 for stimulating discussions on the topics of this paper, and OLAM, Association pour la Recherche Fondamentale, Bruxelles, for support to the conference. Mingyi is supported by CSC scholarship No.2010601003.

\appendix
\section*{Appendix}

\section{Conventions}
We use $A,B,C,\ldots$ for spinor indices in the left-handed representation; $\dot{A},\dot{B},\dot{C},\ldots$  in the right-handed representation; $I,J,K,\ldots$ the Minkowski indices; and $i,j,k,\ldots$ space indices running from $1$ to $3$. A bijection between Minkowski space and spinors is given by
\begin{equation}
M^{A\dot{A}}=\frac{\ii}{\sqrt{2}}M^{I}\sigma^{A\dot{A}}_I,
\end{equation}
where $\s^{A\dot{A}}_I=(1,\vec{\s})$ and $\s^A_{jB}=\s^{A\dot{A}}_j\delta_{B\dot{A}}$ are Pauli matrices. Notice that we are mapping vectors to \emph{anti}-Hermitian matrices consistently with Minkowski metric signature $(-,+,+,+)$. The normalization of the Levi-Civita tensor is $\epsilon^{0123}=1$. We raise and lower spinor indices with
\begin{equation}
\epsilon^{AB}=
\begin{pmatrix}
0 & 1\\
-1 & 0
\end{pmatrix}
=\epsilon_{AB}, \quad \epsilon^{AB}\epsilon_{AC}=\delta^{B}_C, \quad \omega^A=\epsilon^{AB}\omega_B, \quad \omega_A=\epsilon_{BA}\omega^B.
\end{equation}

For the Lorentz algebra, we define
\be
[L^i, L^j] = -\ii \eps^{ijk} L^k, \qquad [L^i, K^j] = -\ii \eps^{ijk} K^k, \qquad [K^i, K^j] = \ii \eps^{ijk} L^k
\ee
in terms of rotations
$L^i\equiv-\frac{1}{2}\epsilon^{0i}{}_{jk}M^{jk}$ and boosts $K^i\equiv M^{0i}$.
We also introduce left-handed ($-$, anti-self-dual) and right-handed ($+$, self-dual) projectors $P_{(\pm)}$, as
\begin{equation}
P_{(\pm)}^{IJ}{}_{KL}=\frac{1}{2}\Big(\delta^{[I}_K\delta^{J]}_L\mp\frac{\ii}{2}\epsilon^{IJ}{}_{KL}\Big),
\end{equation}
and the left-handed generators are defined as
\begin{equation}
\Pi^i :=\ii P_{(-)}^{0i}{}_{IJ}M^{IJ}=\frac{1}{2}(L^i+\ii K^i).
\end{equation}

In general the spinorial form of a bivector is
\be
B^{IJ} = B^{AB} \eps^{\dot A \dot B} + cc,
\ee
where the left-handed and right-handed parts are
\begin{equation}
B^i=P_{(-)}^{0i}{}_{IJ}B^{IJ}=\frac{1}{2}B^{AB}\sigma^i_{AB}, \qquad 
\bar{B}^i=P_{(+)}^{0i}{}_{IJ}B^{IJ}=\frac{1}{2}\bar{B}^{\dot{A}\dot{B}}\bar{\sigma}^i_{\dot{A}\dot{B}}.
\end{equation}
In terms of the self-dual quantities, the Immirzi shift \Ref{MtoB} reads
\begin{equation}
\Pi^i=\frac{\gamma+\ii}{\gamma}B^i, \qquad \Pi^{AB}=-\frac{1}{2}\frac{\gamma+\ii}{\ii\gamma}B^{AB}.
\end{equation}

\section{Null little group}
The group ISO(2), sometimes denoted as E(2), is the symmetry group of two-dimensional Euclidean space $\R^2$. It is not compact, nor semisimple. Its Lie algebra $\mathfrak{iso}(2)$ has  three generators, $J$, $P^1$ and $P^2$, satisfying
\begin{equation}
[J,P^a]=\ii\epsilon^{ab}P^b, \quad [P^a, P^b]=0, \quad (a,b=1,2).
\end{equation}
$J$ is the generator of rotations in $\R^2$, and $P^a$ generate the translations. 

This Lie group appears as the little group of a null direction $N^I$ in Minkowski space, with generators related to the Lorentz generators $M^{IJ}$ by
\begin{equation}
X^{I}=\frac{1}{\sqrt{2}}\epsilon^{I}{}_{JKL}N^JM^{KL}
\end{equation}
Two canonical choices are $N_{\pm}^I = (1,0,0, \pm 1)/\sqrt{2}$. In this two cases, the generators are, 
\begin{align}
& L^3, \qquad P_{+}^1 \equiv P^1 = L^1 - K^2, \qquad P_+^2 \equiv P^2 = L^2+K^1, \\
& L^3, \qquad P_{-}^1 \equiv \wh P^1 = L^1 + K^2, \qquad P_-^2 \equiv \wh P^2 = L^2 - K^1,
\end{align}
and satisfy
\begin{equation}
[L^3,P_{\pm}^a]=\ii\epsilon^{ab} P_{\pm}^b,
\qquad [P^a_{\pm},P^b_{\pm}]=0, \qquad [P^a_{\pm},P^b_{\mp}]=2\ii(\epsilon^{ab} L^3 \pm \d^{ab} K^3).
\end{equation}

On the fundamental representation ${\bf (1/2,0)}$ of $\mathfrak{sl}(2,\mathbb{C})$, the generators are
\begin{equation}
L^3=\frac{1}{2}
\begin{pmatrix}
1 & 0\\
0 & -1
\end{pmatrix}, \quad P^1=
\begin{pmatrix}
0 & -1\\
0 & 0
\end{pmatrix}, \quad P^2=
\begin{pmatrix}
0 & \ii\\
0 & 0
\end{pmatrix},\quad \wh P^1=
\begin{pmatrix}
0 & 0\\
-1 & 0
\end{pmatrix}, \quad \wh P^2=
\begin{pmatrix}
0 & 0\\
-\ii & 0
\end{pmatrix}
\end{equation}
Exponentiating the generators we get the respective group elements, 
\begin{equation}
g^A_{~~B}=
\begin{pmatrix}
\ex^{\frac{\ii}{2}\theta} & -p\\
0 & \ex^{-\frac{\ii}{2}\theta}
\end{pmatrix},\qquad \wh g^A_{~~B}=
\begin{pmatrix}
\ex^{\frac{\ii}{2}\theta} & 0\\
\bar{p} & \ex^{-\frac{\ii}{2}\theta}
\end{pmatrix}, \qquad p:= -p^2+\ii p^1.
\end{equation}

\section{Unitary irreducible representation of $\ISO (2)$ and $\sltc$}

Unitary irreducible representations (irreps) of ISO(2) are complex function $f$ on $\mathbb{C}$, with basis labeled by the eigenvalues $p^a\in \R$ of $P^a$,
\begin{equation}\label{nullbasis}
f_p(z)=\frac{1}{2\pi}\ex^{\frac{\ii}{2}(\bar{z}p+z\bar{p})}, \qquad z=-z^2+\ii z^1,
\qquad p\equiv -p^2+\ii p^1
\end{equation}
\begin{equation}
[P^a\circ f_p](z)=p^af_p(z), \qquad [L^3\circ f_p](z)=(z\p_z-\bar{z}\p_{\bar{z}})f_p(z)
\end{equation}
The basis is orthogonal, 
\begin{equation}
\langle f_p, f_{p'}\rangle=\frac{\ii}{2}\int_{\mathbb{C}}\dd z\wedge \dd \bar{z}~\overline{f_{p}(z)}f_{p'}(z)=\frac{\ii}{8\pi^2}\int_{\C} \dd z\wedge \dd \bar{z} ~\ex^{\frac{\ii}{2}\bar{z}(p'-p)-cc.}=\delta_{\C}(p'-p),
\end{equation} 
and complete,
\begin{equation}
\frac{\ii}{2}\int_{\mathbb{C}}\dd p\wedge \dd \bar{p}~\overline{f_{p}(z)}f_{p}(z')=\frac{\ii}{8\pi^2}\int_{\C} \dd p\wedge \dd \bar{p} ~\ex^{\frac{\ii}{2}\bar{p}(z'-z)-cc.}=\delta_{\C}(z'-z).
\end{equation}
Thanks to these properties, and the induced representations theorem, irreps of $\sltc$ can be spanned by irreps of ISO(2), with a faithful one-to-one map. 

To make the map explicit, recall that irreps of $\sltc$ are built from homogeneous functions on $\mathbb{C}^2$, $f:\mathbb{C}^2\rightarrow\mathbb{C}$. For the principal series, the homogeneity weights can be conveniently parametrized by the pair 
$(\rho, k) \in(\mathbb{R},\mathbb{Z}/2)$ as follows:
\begin{equation}
\forall \lambda\in\mathbb{C}/ \{ 0 \},~ f(\lambda\omega^A)=\lambda^{-k-1+\ii\rho}\bar\lambda^{k-1+\ii\rho}f(\omega^A),
\end{equation} 
and the unitary irrep $D(g)$ of $g^A{}_B=\left(\begin{smallmatrix} a&b\\ c&d \end{smallmatrix}\right)\in\SL(2,\C)$ is given by
\begin{equation}
[D(g)\circ f^{(\rho,k)}](\omega^A)=f^{(\rho,k)}(g^A_{~~B}\omega^B).
\end{equation}  
Then, we define $\om=\om^0/\om^1$, and
\begin{equation}\label{fomegadef}
f^{(\rho,k)}(\omega) := f^{(\rho,k)}\left(\frac{\omega^0}{\omega^1},1\right)=
(\omega^1)^{k+1-\ii\rho}(\bar\omega^{\bar{1}})^{-k+1-\ii\rho}f^{(\rho,k)}(\omega^A).
\end{equation}
By inverting this relation, each homogeneous function $f^{(\rho,k)}(\omega^A)\in H^{(\rho,k)}(\omega^A)$ is uniquely determined by a $f^{(\rho,k)}(\omega)$, and picking in particular the basis \Ref{nullbasis} for the latter, we find
\begin{equation}
f_p^{(\rho,k)}(\omega^A)=(\omega^1)^{-k-1+\ii\rho}(\bar{\omega}^{\bar{1}})^{k-1+\ii\rho}f^{(\rho,k)}_p(\omega)=\frac{1}{2\pi}(\omega^1)^{-k-1+\ii\rho}(\bar{\omega}^{\bar{1}})^{k-1+\ii\rho}\ex^{\frac{\ii}{2}\left(\frac{\bar{\omega}^{\bar{0}}}{\bar{\omega}^{\bar{1}}}p+\frac{\omega^0}{\omega^1}\bar{p}\right)}.
\end{equation}
This defines the null basis for the principal series of $\SL(2,\C)$ irreps.

The $\SL(2,\C)$ action is 
\begin{equation}
[D(g)\circ f^{(\rho,k)}](\omega)=(c\omega+d)^{-k-1+\ii\rho}\overline{(c\omega+d)}^{k-1+\ii\rho}f^{(\rho,k)}\left(\frac{a\omega+b}{c\omega+d}\right),
\end{equation}
and the inner product 
\begin{equation}
\langle f, h\rangle^{(\rho,k)}=\frac{\ii}{2}\int_\mathbb{C} \overline{f^{(\rho,k)}(\omega)} h^{(\rho,k)}(\omega)\dd\omega\wedge\dd\bar\omega=\frac{\ii}{2}\int_{\mathrm{P}\mathbb{C}^2} \overline{f^{(\rho,k)}(\omega^A)} h^{(\rho,k)}(\omega^A)\omega_A\dd\omega^A\wedge\bar{\omega}_{\bar A}\dd\bar\omega^{\bar{A}}.
\end{equation}
In particular,
\begin{equation}
\langle f^{(\rho,k)}_p, f^{(\rho,k)}_{p'}\rangle=\delta_{\C}(p'-p).
\end{equation}


\providecommand{\href}[2]{#2}\begingroup\raggedright\endgroup

\end{document}